\def\theorempostskip#1{}
\def\theorempreskip#1{}

\documentclass[final]{siamart1116}

\hypersetup{pdfpagemode=UseNone}

\usepackage{epstopdf}
\usepackage{braket,amsfonts}
\usepackage[caption=false]{subfig}
\usepackage{amsmath,amssymb}
\usepackage{mathtools}
\usepackage{graphicx}
\usepackage{array}

\Crefname{ALC@unique}{Line}{Lines}

\usepackage{vmargin}
\setpapersize{USletter}
\setmarginsrb{30mm}{19mm}{30mm}{19mm}{10mm}{5mm}{5mm}{5mm}

\DeclareMathOperator{\sech}{sech}

\usepackage{euscript}
\DeclareMathAlphabet{\mathpzc}{OT1}{pzc}{m}{it}

\newcommand{\mynote}[1]{%
  \begin{center}
    \fbox{%
      \begin{minipage}{0.9\linewidth}
        \small\emph{#1}
      \end{minipage}
    }
  \end{center}%
}

\newcommand{\Ht}{H}      % was: z_0
\newcommand{\Rmax}{r_o}  % could be: R
\newcommand{\Rout}{R}    % was: F
\newcommand{\smax}{s_o}  % could be: s_{\text{\emph{max}}}
\newcommand{\Kr}{K_r}    % was: K_{11}
\newcommand{\Kz}{K_z}    % was: K_{22}
\newcommand{\Kmat}{\boldsymbol{\mathsf{K}}}
\newcommand{\myunit}[1]{\mbox{$\mathrm{#1}$}}
\newcommand{\mybunit}[1]{\mbox{$[\mathrm{#1}]$}}
\newcommand{\bigoh}[1]{O\!\left(#1\right)}
\renewcommand{\bigoh}[1]{\EuScript{O}\!\left(#1\right)}
\newcommand{\iimag}{\mathrm{i}}
\newcommand{\defeq}{:=}
\newcommand{\rr}{\tilde{r}}
\newcommand{\Sshift}{\hat{S}}

\newcommand{\leavethisout}[1]{}

\numberwithin{equation}{section}
\numberwithin{theorem}{section}

\title{Asymptotic and numerical analysis of a porous medium model\\
  for transpiration-driven sap flow in trees%
  \thanks{Submitted to the editors DATE.  \funding{This work was funded
      in part by a Discovery Grant from the Natural Sciences and
      Engineering Research Council of Canada and a grant from
      the North American Maple Syrup Council Research Fund.}}}

\author{%
  Bebart Maisar Janbek\thanks{Department of Mathematics, Simon Fraser
    University, 8888 University Drive, Burnaby, BC, V5A 1S6, Canada
  (\email{bjanbek@sfu.ca}, \email{jstockie@sfu.ca}).}
  \and
  John M. Stockie\footnotemark[2]
}

\headers{Porous medium model for sap flow in trees}{B.\ M.\ Janbek and
  J.~M.\ Stockie} 

\begin{document}

\maketitle
  
\begin{abstract}
  We develop a 3D porous medium model for sap flow within a tree stem,
  which consists of a nonlinear parabolic partial differential equation
  with a suitable transpiration source term.  Using an asymptotic
  analysis, we derive approximate series solutions for the liquid
  saturation and sap velocity for a general class of coefficient
  functions.  Several important non-dimensional parameters are
  identified that can be used to characterize various flow regimes.  We
  investigate the relative importance of stem aspect ratio versus
  anisotropy in the sapwood hydraulic conductivity, and how these two
  effects impact the radial and vertical components of sap velocity.
  The analytical results are validated by means of a second-order finite
  volume discretization of the governing equations, and comparisons are
  drawn to experimental results on Norway spruce trees.
\end{abstract}
  


%\tableofcontents

%%%%%%%%%%%%%%%%%%%%%%%%%%%%%%%%%%%%%%%%%%%%%%%%%%%%%%%%%%%%%%%%%%%%%%%%%%%
%%%%%%%%%%%%%%%%%%%%%%%%%%%%%%%%%%%%%%%%%%%%%%%%%%%%%%%%%%%%%%%%%%%%%%%%%%%
%%%%%%%%%%%%%%%%%%%%%%%%%%%%%%%%%%%%%%%%%%%%%%%%%%%%%%%%%%%%%%%%%%%%%%%%%%%
%%%%%%%%%%%%%%%%%%%%%%%%%%%%%%%%%%%%%%%%%%%%%%%%%%%%%%%%%%%%%%%%%%%%%%%%%%%
\section{Introduction}
\label{sec:intro}

The phenomenon of sap transport in trees has challenged 
plant physiologists for over a century~\cite{dixon-joly-1894}.  The
remarkable ability of trees to draw water upward from roots to leaves
through heights of 100 metres or more is still not completely 
understood and still generates a great deal of interest in the
biophysics and plant physiology communities \cite{angeles-etal-2004,
  brown2013theory, jensen-etal-2016}.  The driving process behind sap
flow is \emph{transpiration}, which can be briefly explained as
follows~\cite{taiz2002plant}.  Water is lost from the leaves as a
consequence of the shared path between inward diffusion of carbon
dioxide (the essential precursor for photosynthesis) from the ambient
air to the leaves, and the combined outward evaporation/diffusion of
water due to the vapor pressure deficit.  As water evaporates from the
leaves, air-water surface tension is generated within the hydrophilic
leaf cell interstices and is subsequently transmitted through the sap to
conductive wood cells (known as \emph{sapwood} or \emph{xylem}) making
up the stem and branches.  The resultant negative sap pressure (measured
relative to atmospheric) generates the driving force necessary to draw
water from the roots to the leaves~\cite{pickard1981ascent}.  This
transpiration-driven flow is a maximum during the day when there is a
high evaporative demand, and drops to a minimum overnight when
photosynthesis halts and the internal water storage within the tree is
replenished \cite{zimmermann1983xylem}.

The physical structure of wood plays an essential role in sap transport.
Within coniferous (or softwood) tree species that are the focus of this
study the primary conductive elements in sapwood are the tracheids,
which are elongated and vertically-oriented dead wood cells having rigid
lignified cellulose walls enclosing an empty lumen.  Adjacent tracheids
are hydraulically connected through paired pits, which are pores that
permit flow of liquid sap but are small enough to prevent passage of any
air bubbles that might be formed within the tracheids.  When the sapwood
is viewed as a porous medium, the combination of pit distribution and
tracheid orientation engenders a high degree of anisotropy in hydraulic
conductivity, which can be several orders of magnitude larger in the
vertical direction than the radial~\cite{comstock-1970}.  Deciduous (or
hardwood) tree species differ from conifers in that sapwood contains an
extra class of wood cells called \emph{vessels} having a much larger
diameter and with greater permeability to flow, although the wood
structure and transport properties remain otherwise similar.

Because typical pressures within a tree stem exceed the saturated vapor
pressure, gases such as air and carbon dioxide are known to exist in a
dissolved state within the sap.  Under conditions of extreme dryness or
freezing, pressure can fall enough to cause dissolved gases to cavitate
and form air bubbles inside tracheids or
vessels~\cite{zimmermann1983xylem}.  This process is known as
\emph{embolism} and can lead to blockage of affected conduits that
prevents them from conducting sap.  Trees have an amazing microstructure
that is capable of bypassing and even eliminating such
embolisms~\cite{cruiziat2002hydraulic}, but the precise mechanisms for
embolism formation and recovery are still not completely
understood~\cite{brodersen2013maintenance}.  To avoid having to model
the added complexities of embolism formation, we assume in this study
that the trees operate under ``normal conditions'' in which the sapwood
remains close to full saturation so that no new embolus forms, and
neither does embolism recovery play a significant role.

Numerous mathematical have been developed for studying the flow of sap
within conductive sapwood.  One of the most popular approaches is an
electric circuit analogy, in which the porous flow elements are
characterized by a flow resistance (which is inversely proportional to
conductivity), combined with capacitive elements that capture storage of
sap (and subsequent time lags) within the roots, leaves and various
cells making up the stems and branches~\cite{hunt1987non}.  The earliest
circuit models included only resistive elements and hence failed to
capture the dynamic nature of the flow~\cite{van1948water}, whereas more
recent models include water storage effects and hence capture observed
lags between transpiration flux and sap flow~\cite{phillips1997time}.
Some authors have developed even more detailed models that capture the
branching structure of the tree~\cite{tyree1988dynamic}, or incorporate
the added dynamic effects of radial stem
growth~\cite{steppe2006mathematical} by connecting resistor-capacitor
elements in a more complex branching structure.  A major drawback of
these circuit models is that the parameters have no direct
correspondence in the context of an actual porous medium flow, and the
resistance and capacitance are treated instead as fitting constants that
are matched to experiments rather than being directly measured.
Furthermore, the circuit analogue is constructed out of a network of
resistors and capacitors connected in series, for which the resulting
flow depends sensitively on the actual discretization used.  Finally,
the circuit parameters are usually treated as constants although the
transport properties of actual sapwood change in time owing to the local
saturation state. Having said that, circuit models are still applied
widely in the tree physiology literature because of their simplicity and
straightforward algebraic structure.
 
Another type of model that overcomes many of these deficiencies is the
class of porous medium models.  The study of unsaturated porous flow in
porous media is very well developed~\cite{klute1952numerical},
especially in the context of groundwater transport in
soils~\cite{cowan1965transport}.  Sap flow is especially suitable for
treatment using continuum porous medium models because of the simple
repeating microstructure of wood.  In this class of models, sap flow is
driven by pressure gradients according to Darcy's law, and the governing
equations consist of nonlinear partial differential equations (PDEs)
that capture spatial and temporal variations in variables such as water
content (saturation) and pressure. Chuang et al.~\cite{chuang2006porous}
developed a simple 1D porous medium model for transpiration driven flow
in a conifer stem, and used numerical simulations to fit their results
to experimentally measured sap fluxes. Bohrer et
al.~\cite{bohrer2005finite} extended these results to include the effect
of complex branching structure within the crown \cite{chuang2006porous}.
Aumann and Ford \cite{aumann2002modeling, aumann2002parameterizing}
applied a different approach focused on the wood microstructure in
Douglas fir by developing a detailed model for transport within a
tracheid network including water-air interface dynamics and many
microstructural parameters.  Such a model is useful for uncovering
detailed aspects of flow within individual wood cells, but it is not
very practical for use at the scale of an entire tree.  A number of
other related PDE models have been also developed~\cite{fruh-kurth-1999,
  kumagai-2001, peramaki-vesala-nikinmaa-2005, reid-etal-2005}, but many
are one-dimensional and so ignore effects such as radial variations
within the stem, not to mention that to date little mathematical
analysis has been done to determine the character of the solutions.

The main goal of this paper is to extend the 1D porous medium model
from~\cite{chuang2006porous} to a more realistic 3D cylindrical
(rotationally symmetric) model of a tree trunk, with an imposed
transpiration flux distributed along the outer surface.  Along with
realistic coefficient functions fit to data on Norway spruce, this model
will permit study of the radial flow patters that develop with the stem.
We will then develop an approximate analytical solution with the aid of
asymptotic analysis, identifying the different parameter regimes where
anisotropy begins to dominate which provides an alternate explanation
for the observed radial variation in the vertical velocity
\cite{james2003axial, poyatos-etal-2007} that isn't due to loss of
hydraulic conductivity.  We study the relative importance of gravity and
transpiration as well as investigating the nature and relative
importance of the radial/vertical sap fluxes, which is significant in
light of the recent experimental advances that permit measurements of
the relatively small radial sap velocity
components~\cite{domec2006transpiration, testi2009new}.  We further
demonstrate how temporal and spatially localized disturbances in
saturation propagate along the tree.  Throughout, we use a finite volume
discretization of the governing nonlinear PDE in order to solve the
problem numerically and validate the asymptotic results.

\section{Model for 3D Axisymmetric Flow in a Tapered Cylindrical
  Annulus} 
\label{sec:model}

Our work is based on the one-dimensional sap flow model of Chuang et
al.~\cite{chuang2006porous}, who considered only vertical variations
within a tapered cylindrical tree stem.  We extend their model to a 3D
axisymmetric stem geometry by incorporating the effect of radial
variations as well as a core heartwood region that is impermeable to
flow.  Except for slight changes in notation and the need for
several additional boundary conditions, the model is much the same as
the 1D analogue.  A description of all parameters and solution-dependent
coefficient functions is provided in \cref{sec:chuang-params}, which are
obtained from the literature or fit to experimental 
measurements from Norway spruce (\emph{Picea abies}) provided
in~\cite{chuang2006porous}.

\subsection{Governing Equations and Boundary Conditions}
\label{sec:model-3d}

\begin{figure}[tbhp]
  \centering\footnotesize
  \begin{tabular}{l@{}lll@{}l}
    (a) &&& (b) \\[-0.7cm]
    & \includegraphics[trim=1.5cm 0 0 0,clip,width=0.23\textwidth]{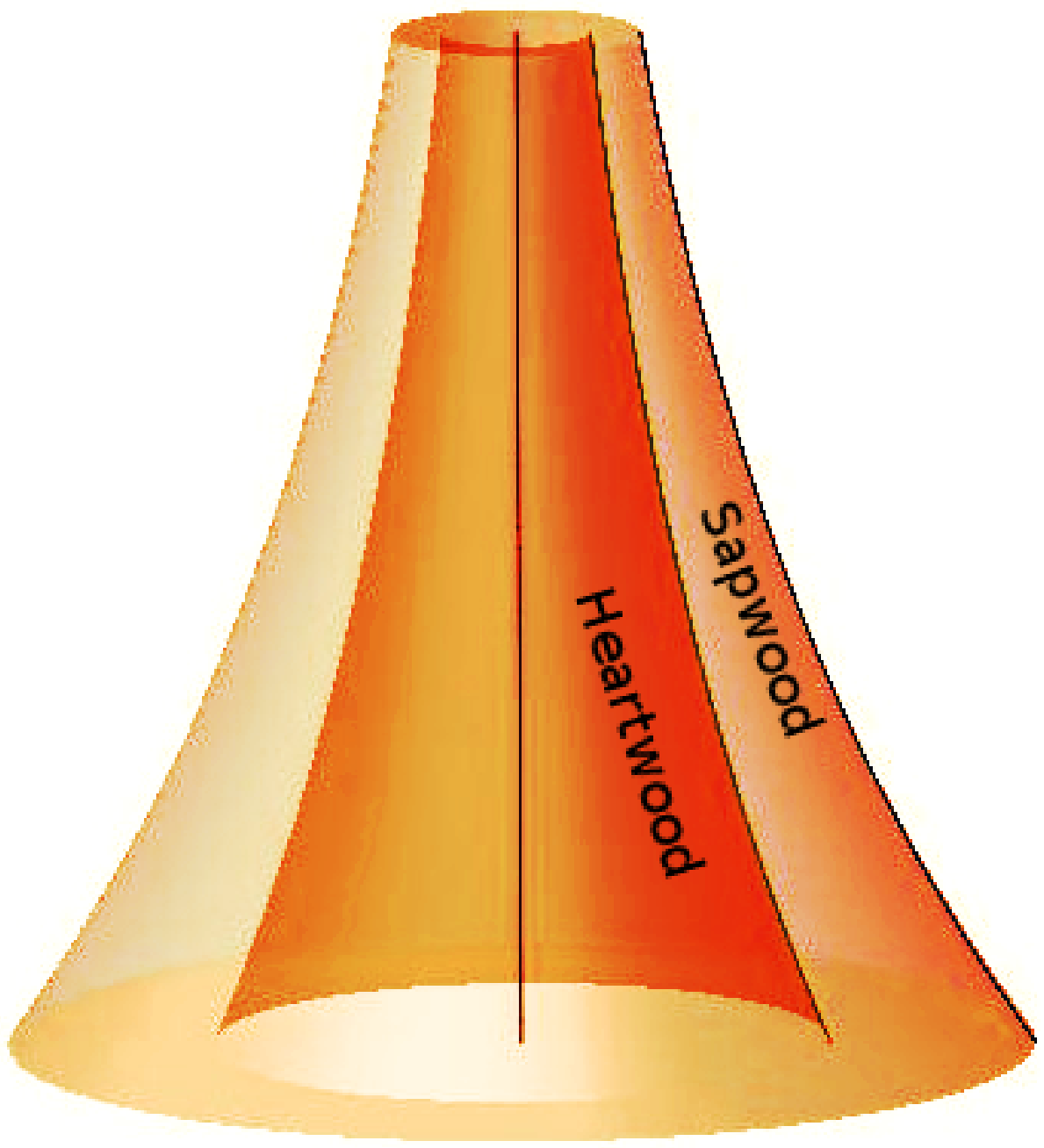} 
    &&
    & \includegraphics[trim=3cm 0 0 0,clip,width=0.54\textwidth]{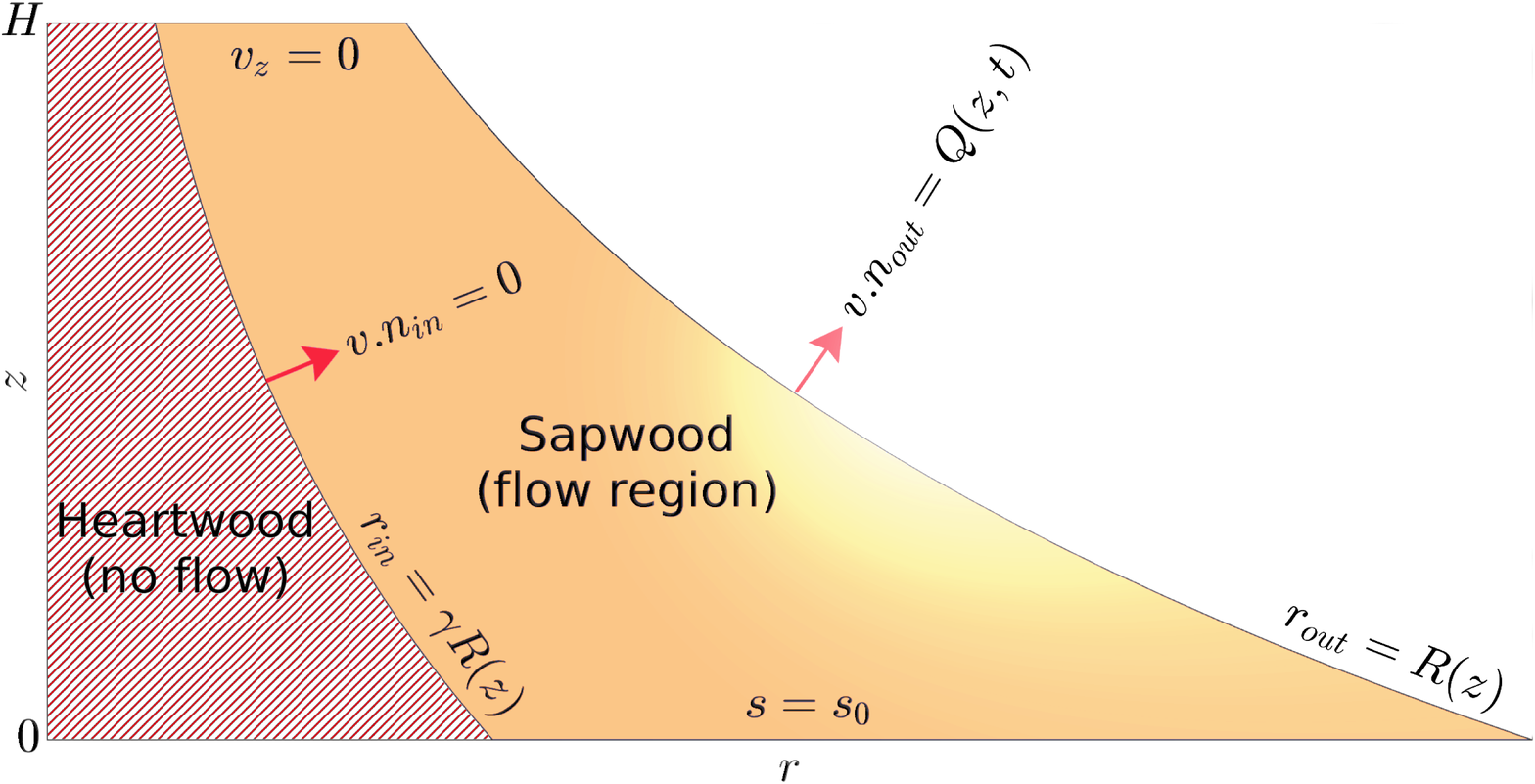}%or 6
  \end{tabular}
  \caption{(a) Tree stem and heartwood regions, both having the shape of
    tapered circular cylinders.  (b) A vertical cross-section depicting
    the annular sapwood domain and boundary conditions along each
    boundary segment.}
  \label{fig:BCsSap}
\end{figure}

The Norway spruce is a conifer species whose stem is well-approximated
by a right circular cylinder that tapers from base to crown.  For a tree
of height $H$ (in units of \myunit{m}) the outer stem radius may be
written as $r=\Rout(z)$ \mybunit{m}, where the axial coordinate $z$
satisfies $0\leqslant z\leqslant \Ht$ and $\Rout(z)$ is a decreasing
function of $z$ (which will be specified later in
\cref{sec:chuang-params}).  We extend the 1D model
from~\cite{chuang2006porous} by taking a more realistic geometry
pictured in \cref{fig:BCsSap}a where the stem consists of an outer layer
of conducting sapwood that surrounds a core region of non-conducting
heartwood.  The heartwood is assumed to take up some 
fraction $0<\gamma<1$ of the stem cross-section corresponding to
$0\leqslant r\leqslant \gamma \Rout(z)$.  We are therefore concerned
with capturing sap transport within the outer conductive portion which
has the shape of a tapered cylindrical annulus defined by 
\begin{gather}
  \gamma \Rout(z) \leqslant r \leqslant \Rout(z)
  \qquad \text{and} \qquad 
  0 \leqslant z \leqslant \Ht.
  \label{eq:domain}
\end{gather}
In the absence of any directional forcing around the stem, we can suppose
rotational symmetry and neglect any dependence on the polar angle.
% However, we will also consider the case $\gamma=0$ where the entire stem
% conducts sap and which is relevant to younger saplings that haven't yet
% developed a significant heartwood region.

The sapwood is treated as a variably-saturated medium whose porous
structure contains a mixture of two phases: liquid (sap) and gas (mostly
air).  The saturation or local pore volume fraction containing liquid is
denoted by $s(r,z,t)$, and depends on location and time $t$ \mybunit{s}.
Enforcing conservation of liquid yields the continuity equation
\begin{gather}
  \frac{\partial s}{\partial t} + \nabla \cdot \vec{v} = 0, 
  \label{eq:continuity-3d}
\end{gather}
where the sap velocity $\vec{v}(r,z,t)$ \mybunit{m/s} obeys Darcy's law
\begin{gather*}
  \vec{v} = - \Kmat \, \nabla (z + \psi(s)).
  \label{eq:darcy-vec}
\end{gather*}
Here $\psi(s)$ \mybunit{m} is the hydrostatic pressure head, which is a
saturation-dependent function that is specified later in
\cref{sec:chuang-params}.  The hydraulic conductivity tensor $\Kmat$
\mybunit{m/s} depends in general on both the location within the tree
stem (due to spatial variations in pore structure) and the local
saturation state.  The porous structure of wood is highly anisotropic
since it is composed of dead, hollow, elongated cells that are
directed vertically within the xylem; in conifers, these cells are
tracheids, whereas in deciduous trees the primary route for sap
transport is through much larger cells called vessels.  Microscopic
``pits'' or pores connect adjacent wood cells in the radial direction,
as well as a much smaller number of radially-directed ``ray cells'';
however, these structures are much less permeable to flow than the
vertically-oriented tracheids, so it is reasonable to assume that the
anisotropic hydraulic conductivity has the form of a diagonal tensor
\begin{gather}
  \Kmat =
  \begin{bmatrix}
    \Kr(r,z) & 0  \\
    0      & \Kz(r,z)
  \end{bmatrix} . 
  \label{eq:Ktensor}
\end{gather}
Note that $\Kmat$  depends on position but not on
saturation, which we justify based on the earlier assumptions that
sapwood remains nearly saturated and no significant build-up of gas
occurs owing to absence of embolism formation.  This should be
contrasted with Chuang et al.~\cite{chuang2006porous} who took the
conductivity to be a function of saturation; however, since we are
primarily concerned with trees near the fully saturated state where the
hydraulic conductivity is relatively insensitive to changes in $s$,
taking $\Kmat$ to be independent of $s$ is a reasonable approximation
(to be discussed more detail in the next section).

In most trees, the dominant vertical orientation of conductive cells
induces anisotropy in $\Kmat$ that ranges from moderate to extreme.
Therefore, the analytical derivations in this paper focus on two
limiting cases: an isotropic conductivity with $\Kr \sim \Kz$, for which
a general series solution is relatively straightforward to derive; and
high anisotropy with $\Kr \ll \Kz$, which can be treated analytically
under certain restrictions.  In either case, numerical simulations are
used to study the solution over the entire range of anisotropy.

Using the above definitions, the velocity can be expressed in
cylindrical coordinates as
\begin{gather}
  \vec{v} = (v_r, \; v_z) = \left(
    -\Kr \, \frac{\partial\psi}{\partial r}, \;\;
    -\Kz \left(1 + \frac{\partial\psi}{\partial z} \right)
  \right).
  \label{eq:darcy-2d}
\end{gather}
After substituting these two expressions into the continuity equation
\cref{eq:continuity-3d}, we obtain the following nonlinear parabolic
PDE for saturation
\begin{gather}
  \frac{\partial s}{\partial t} = \frac{1}{r} \frac{\partial}{\partial
    r} \left( r \Kr \psi^\prime \, \frac{\partial s}{\partial r} \right) +
  \frac{\partial}{\partial z} \left( \Kz \psi^\prime \, \frac{\partial
      s}{\partial z} \right) + \frac{\partial \Kz}{\partial z} .
  \label{eq:Maindim}
\end{gather}

Moving on to the boundary conditions, no sap flows through the top of
the stem so that  
\begin{gather}
  v_z(r,\Ht,t) = 0,
  \label{eq:sapBCtop}
\end{gather}
whereas the base of the stem (at the roots) is assumed to be fully
saturated with
\begin{gather}
  s(r,0,t) = \smax.
  \label{eq:sapBCbottom}
\end{gather}
On the inner sapwood boundary, a zero flux is imposed
\begin{gather}
  \vec{v} \cdot \hat{n}|_{r=\gamma \Rout(z)} = 0, 
  \label{eq:sapBCleft}
\end{gather}
where $\hat{n}(z)=(1,-\Rout^\prime)/\sqrt{1+(\Rout^\prime)^2}$ is the
unit outward-pointing surface normal.  This last condition reflects the
fact that no sap exchange occurs with the non-conducting heartwood (and
reduces to a simple radial symmetry condition in the case $\gamma=0$).

The final boundary condition on the outer stem surface derives from the
transpiration flux, which is the process whereby water is drawn upward
from the roots, through the stem and branches to the leaves (or
needles).  Transpiration is driven by leaf evaporation and the
subsequent vapor pressure deficit between the leaf interior and the
surrounding atmosphere. A particular feature of species like Norway
spruce is that branches are distributed densely along the entire stem so
that the transpiration flux can be specified as a corresponding
distribution in the axial direction (this should be contrasted with many
deciduous species in which branches are concentrated within the crown at
the top of a long, bare trunk).  Therefore, we impose a sap outflow due
to branches distributed continuously along the stem by means of another
flux boundary condition
\begin{gather}
  \vec{v}\cdot\hat{n}|_{r=\Rout(z)} = Q(z,t). %=f(z)E(t)
  \label{eq:sapBCRout}
\end{gather}
The transpiration flux $Q$ \mybunit{m^3/m^2\,s} depends on time owing to
diurnal variations in transpiration, as well depending on the branch
distribution along the stem, and we provide a specific functional form
for $Q(z,t)$ in the next section.  A primary difference from the 1D
model in \cite{chuang2006porous} is that the transpiration flux was
incorporated as a source term in the saturation
equation~\cref{eq:Maindim} rather than as a boundary condition.  A
graphical summary of the geometry and boundary conditions is provided in
\cref{fig:BCsSap}b.

\subsection{Parameters and Coefficient Functions}
\label{sec:chuang-params}

We now provide parameter estimates and functional forms for the model
coefficients, which are chosen to match as closely as possible the data
for Norway spruce provided in~\cite{chuang2006porous}.  All symbols are
listed in \cref{tab:tableValues} along with units and values.  We
emphasize that neither the model nor the asymptotic analysis in the
following sections is restricted to a particular species of tree, and
that the specific functional forms introduced here are not critical for
the analysis.  Rather, we exploit these parameters as a convenient
illustration of our 3D model, which has the added benefit of allowing us
to draw a concrete comparison with the results
in~\cite{chuang2006porous}.

\newcommand{\mymcc}[1]{\multicolumn{1}{l}{#1}}
\begin{table}[tbhp]
  \footnotesize
  \centering
  \setlength{\extrarowheight}{0.07cm}
  \caption{Variables and parameters for the 3D sap flow model.  Listed
    values are for the ``base case'' used in the asymptotic analysis,
    while parameters from additional simulations are given in
    parentheses.}  
  \label{tab:tableValues}
  \begin{tabular}{|c|l|c|c|}\hline
    % \multicolumn{4}{|c|}{nomenclature}   \\\hline
    Symbol & Description & Units & Value or Formula\\\hline 
    \multicolumn{4}{|l|}{Independent and dependent variables:}\\\hline
    $r$      & radial coordinate             & \myunit{m}   & \\
    $z$      & vertical coordinate           & \myunit{m}   & \\
    $t$      & time                          & \myunit{s}   & \\
    $s(r,z,t)$  & sap volume fraction or saturation & --    & \\
    $S(r,z,t)$  & saturation deficit         & --           & \\
    $v_r(r,z,t)$& radial sap velocity        & \myunit{m/s} & \\
    $v_z(r,z,t)$& vertical sap velocity      & \myunit{m/s} & \\\hline
    \multicolumn{4}{|l|}{Solution-dependent functions:}\\\hline
    $E(t)$   & transpiration rate per unit leaf area  & \myunit{m^3/m^2\, s} & \cref{eq:Et} \\
    $f(z)$   & combined leaf area and shading effects & --   & \cref{eq:fz} \\
    $\Kmat$  & hydraulic conductivity tensor & \myunit{m/s}  & \cref{eq:Ktensor} \\
    $K_{r,z}$& hydraulic conductivity components & \myunit{m/s}  & \cref{eq:Krz} \\
    $\ell(z)$& leaf area per unit height     & \myunit{m^2/m}& \cref{eq:ellz} \\
    $\Rout(z)$& outer stem radius            & \myunit{m}    & \cref{eq:Rout} \\
    $\lambda(z)$ & sunlight shading effect   & -- & \cref{eq:lambdaz} \\
    $\psi(s)$& capillary pressure head       & \myunit{m}    & \cref{eq:psis} \\\hline
    \multicolumn{4}{|l|}{Physical (dimensional) parameters:}\\\hline
    $E_o$    & transpiration flux amplitude  & \myunit{m/s} & $1\times10^{-9}$~($3.94\times 10^{-8}$)\\
    % $\epsilon_o$& transpiration rate       & \myunit{m/s}  & $1.125\times 10^{-8}$ \\  
    % $g$    & gravitational acceleration    & \myunit{m/s^2}& 9.8 \\
    $\Ht$    & stem height                   & \myunit{m}    & 6.7  \\
    $K_o$    & maximum hydraulic conductivity& \myunit{m/s}  & $5.36\times 10^{-7}$   \\
    $\ell_o$ & leaf specific area            & \myunit{m}    & 15.3 \\
    % $p_o$  & fitting parameter for $K$     & \myunit{m}    & $694$   \\
    $\Rmax$  & maximum tree radius (at base) & \myunit{m}    & 0.0645 \\
    % $\rho$ & sap density                   & \myunit{kg/m^3} & 1000  \\
    $\tau$   & number of seconds per day     & \myunit{s}    & $8.64\times 10^4$ \\
    $\psi_o$ & scaling constant in $\psi(s)$ & \myunit{m}    & $2.93\times 10^5$  \\\hline
    \multicolumn{4}{|l|}{Dimensionless parameters:}\\\hline
    $f_o$    & \mymcc{maximum of $f(z)$}                     & & 2.6 \\
    $n$      & \mymcc{capillary pressure exponent}           & & 400 \\
    $\smax$  & \mymcc{maximum saturation}                    & & 0.574  \\
    $\alpha$ & \mymcc{exponential stem taper rate}           & & 1.42 \\
    % $\beta$& \mymcc{fitting parameter for $k(s)$}          & & 3.5 \\
    $\gamma$ & \mymcc{heartwood fraction (inner : outer radius ratio)} & & 0~(0--0.75) \\
    $\delta$ & \mymcc{relative change from max.\ saturation} & & 0.01 \\
    $\zeta$  & \mymcc{stem aspect ratio\ $=\Rmax/\Ht$}       & & 0.00963 \\
    $\eta$   & \mymcc{time parameter\ $=(2\pi n\smax\Ht^2)/(\tau\psi_o K_o)$} & & $4.77$ \\
    $\kappa$ & \mymcc{conductivity ratio\ $=\Kr/\Kz$}        & & $10^{-4}$--$10^{-2}$ \\
    $\mu$    & \mymcc{gravity parameter\ $=n\Ht/\psi_o$}     & & 0.00915 \\
    $\phi$   & \mymcc{transpiration parameter\ $=2 f_oE_o\mu/K_o\zeta$} & & 0.00920~(0.363)\\
    \hline
  \end{tabular}
\end{table}

We first specify the form of the outer tree stem radius
that tapers with height according to
\begin{gather}
  \Rout(z) = \Rmax \exp\left(-\frac{\alpha z}{\Ht} \right), 
  \label{eq:Rout}
\end{gather} 
where $\alpha$ controls the rate of taper from roots to crown.  This
choice of exponential function was motivated in~\cite{chuang2006porous}
for reasons of mathematical convenience.  However, there is an extensive
literature on more complicated diameter-versus-height
relationships~\cite{li-weiskittel-2010, niklas-1995}, from which we
observe that many coniferous and deciduous tree species have a small
enough taper rate that such an exponential function provides a
reasonable approximation of stem shape.  However, we still develop the
majority of our analytical results for the general function $\Rout(z)$
and impose \cref{eq:Rout} when we need to exploit additional
simplifications in \cref{sec:special-case}, and in numerical simulations
when a specific form of $\Rout(z)$ is required.

Another important geometric parameter is the heartwood ratio $\gamma$
that determines the thickness of the annular sapwood region.  For young
trees that have not yet developed a well-defined heartwood region,
taking $\gamma=0$ is a reasonable choice.  For spruce trees of the age
and height considered in \cite{chuang2006porous} the typical heartwood
fraction is $\gamma\approx 0.5$~\cite{sellin-1994}, while for other
species $\gamma$ can be as high as 0.75~\cite{wullschleger-king-2000}.
For most of this paper including the asymptotic developments we assume
that $\gamma=0$, although we explain later how our results can be
extended to the case when the heartwood fraction is much larger.

Based on our earlier assumption that the hydraulic conductivity is a
diagonal tensor with entries that depend only on position, we take the
axial and radial conductivities to have the form
\begin{gather}
  \Kz = K_o K^*(r,z) \qquad \text{and} \qquad
  \Kr = \kappa K_o K^*(r,z), 
  \label{eq:Krz}
\end{gather}
where $K_o$ \mybunit{m/s} is a constant equal to the maximum value of
axial hydraulic conductivity, and $K^*$ is a dimensionless function that
is strictly positive.  The dimensionless factor $\kappa$ is the ratio of
radial to axial conductivity that captures the degree of anisotropy in
the sapwood.  Typical values of $\kappa$ lie between $10^{-4}$ to
$10^{-2}$~\cite{comstock-1970, redman-etal-2012}, although we will also
consider the case when $\kappa\sim{1}$ (where we use $\sim$ to denote
asymptotic equivalence in which the two quantities have the same order
of magnitude).

\leavethisout{
  $K(s)$ is a (dimensionless) Weibull-type vulnerability curve
  \begin{gather}
    K^*(s) = \exp\left[ -\left(
        \frac{-\psi(s)}{p_o} \right)^\beta \right] 
    % \frac{\rho g\psi(s)}{p_o} \right)^\beta \right] 
    \label{eq:Ks} 
  \end{gather}
  that is commonly used in the tree hydraulics literature to represent
  saturation-dependent conductivity in sapwood~\cite{sperry-etal-1998},
  with $p_o$ \mybunit{m} and $\beta$ being fitting parameters.  
}
% Also: \cite{agoua-perre-2010} 
% 
% The parameter $\rho$ \mybunit{kg/m^3} is the sap density,
% $g$ \mybunit{m/s^2} is the gravitational acceleration.

Many models of variably-saturated porous media
specify the hydraulic conductivity as a function of saturation; indeed,
Chuang et al.~\cite{chuang2006porous} imposed a Weibull-type function of
the form $K^*(s)=\exp(-a|\psi(s)|^b)$. 
% a = 1/694 = 0.00144, b = 3.5
However, their simulations remained within 20\%\ of the fully saturated
state $s\approx\smax$, for which $K^*(s)$ has a nearly linear dependence
on $s$ with a very small (negative) slope.  The reason for this
behaviour is that the primary cause of hydraulic conductivity loss in
trees is embolism formation, which only occurs when saturation drops
sufficiently below $\smax$.  Since we expect that variations in
saturation remain relatively small, it is reasonable to approximate
variations in conductivity using a simpler spatially dependent function
$K^*(r,z)$ for which we derive most of our asymptotic results (and which
avoids a nonlinear $K^*$).  After investigating the asymptotic solution
for spatially-dependent conductivity, we then consider in
\cref{sec:special-case} the special case of constant conductivity and a
periodic transpiration rate, which allows us to derive a simpler
closed--form solution for the leading order term.

The hydraulic pressure head is taken to depend on saturation according to
\begin{gather}
  \psi(s) = \psi_o \left[ 1 - \left(\frac{\smax}{s}\right)^{1/n}
  \right],   
  \label{eq:psis}
\end{gather}
where $\psi_o$ and $n$ are fitting parameters.  This 
is similar to the van Genuchten model commonly used for
capillary pressure in soil and rock~\cite{szymkiewicz-2013}, and 
also applied to drying of lumber~\cite{kang-chung-2009}.  Note that
the head is a negative quantity because $s<\smax$, reflecting the
understanding in the sap hydraulics literature that sap within a tree
stem is under tension.
% (and also explains the extra negative sign in \cref{eq:Ks}).
Our asymptotic derivation is not specific to this or
any other particular form of the capillary pressure function, but does
rely on two essential features: namely, that $\psi(s)$ is a smooth and
monotone increasing function in a neighbourhood of $s=\smax$.
% Smooth: abrupt jumps and hysteresis may occur only due to embolism,
% which we assume does not happen due to the small saturation variations
%
% Concave downwards: as it becomes ever harder to extract moisture from
% the tissue surrounding the xylem, as water recedes into the
% interstices of cell walls, and the cell palsma becomes more
% concentrated (osmotic pressure).  

\leavethisout{
\mynote{I was concerned about this parameterization, and I was
  originally afraid that Chuang made a mistake!  But not anymore, thanks
  to our recent discussions. The value of $n=400$ ($P=400$ in Chuang's
  notation) seemed to be inconsistent with the porous media literature.
  The equivalent van Genuchten expression is
  \begin{gather*}
    \psi(s) = \psi_o \left[ 1 - \left(\frac{\smax}{s}\right)^{1/n}
    \right]^{1/m}
  \end{gather*}
  where $n=1-\frac{1}{m}$, so it seemed that \cref{eq:psis} corresponds
  to $m\to 1$, or $n\to 0$ (certainly not 400!).  Any studies I have
  read on soil, rock, or concrete report $0\leqslant n \leqslant 1$.
  Furthermore, Kang and Chung~\cite{kang-chung-2009} report $n\in[0.10,
  0.56]$ for a variety of tree species, both softwoods and hardwoods.
  But a plot of $\psi(s)$ shows that the curves have a similar shape, so
  all we need to do is add a brief discussion of this point.  Finally, a
  notational nit-pick: the standard notation in the porous media
  literature has the meanings for exponents $n$ and $m$ reversed, and so
  it might be worthwhile to switch to using $m$ instead of $n$
  in~\cref{eq:psis}.}  }

The final ingredient in the model specification is the transpiration
source term, which we assume takes the separable form
\begin{subequations}
  \label{eq:Qsource}
\begin{gather}
  Q(z,t) = f(z) \, E(t),  
  \label{eq:source-sep}
\end{gather}
where $E(t)$ captures time variations throughout the daily transpiration
cycle while $f(z)$ embodies changes with height.  In general, our only
requirement on the source term is that the time-dependent factor $E(t)$
is periodic and can be expressed as a Fourier series; however, for
illustration purposes we choose particular forms for both functions that
approximate the experimental data provided in~\cite{chuang2006porous}.
For the time-dependent factor we take a periodic function
\begin{gather}
  % E(t) = \frac{2E_o}{3} \left[ 1-\cos\left(\frac{2\pi t}{\tau}\right) \right]^2 , 
  E(t) = E_o \Re \big[ 1 + 
    d_1 \exp \left( {2\pi\iimag t}/{\tau} \right) + 
    d_2 \exp \left( {4\pi\iimag t}/{\tau} \right) \big] 
  \label{eq:Et}
\end{gather}
% d2=-0.9118 - 0.0494i; 
% d3=-0.0446 + 0.1861i;
consisting of a three-term Fourier expansion, where time $t$ is measured
from midnight on the first day, $\tau=86,400$~\myunit{s} is the diurnal
period, and $E_o = 3.94\times 10^{-8}$~\myunit{m/s} is the amplitude of
the transpiration flux.  The complex fitting parameters $d_1$ and $d_2$
are obtained by digitizing 36 data points from \cite[Fig.~7
(bottom,~CC)]{chuang2006porous} and then using a discrete Fourier
transform to extract the first three Fourier coefficients, which are
sufficient to obtain a smooth approximation of the original data.

The height-dependent transpiration factor is decomposed as
\begin{gather}
  f(z) = \frac{\ell(z) \lambda(z)}{2\pi \Rout(z)}, 
  \label{eq:fz}
\end{gather}
where $2\pi\Rout(z)$ is a geometric scaling factor (equal to stem
perimeter) and $\ell(z)$ is leaf area density (units of \myunit{m^2/m})
that captures the impact of sun exposure on transpiration and is given
in~\cite[Fig.~4]{chuang2006porous} as
\begin{gather}
  \ell(z) = \ell_o\sech^2\left(\frac{6z}{\Ht} - 2.4\right) .
  \label{eq:ellz}
\end{gather}
The remaining factor $\lambda(z)$ is a dimensionless quantity called
transpiration flux density that captures shading effects due to
branches/leaves located above a given height.  Although $\lambda(z)$ was
not provided in~\cite{chuang2006porous}, it can be approximated using
experimental data in~\cite[Fig.~5]{chuang2006porous} as
\begin{gather}
  \lambda(z) = \frac{1}{\pi} \arctan\left( \frac{63 z}{\Ht} - 50 \right)
  + 0.53.
  \label{eq:lambdaz}
\end{gather}
\end{subequations}

To summarize, the 3D sap flow model corresponds to solving
\crefrange{eq:Maindim}{eq:sapBCRout} along with a suitable initial
condition on saturation and the function definitions in
\cref{eq:Rout}--\cref{eq:Qsource}.
\leavethisout{
  The corresponding 1D model
  consists of \crefrange{eq:sat-1d}{eq:BCH-1d} along with the same
  function definitions, except that the source term must be rescaled using
  \cref{eq:source-convert}.
}%

%%%%%%%%%%%%%%%%%%%%%%%%%%%%%%%%%%%%%%%%%%%%%%%%%%%%%%%%%%%%%%%%%%%%%%%%%%%
%%%%%%%%%%%%%%%%%%%%%%%%%%%%%%%%%%%%%%%%%%%%%%%%%%%%%%%%%%%%%%%%%%%%%%%%%%%
%%%%%%%%%%%%%%%%%%%%%%%%%%%%%%%%%%%%%%%%%%%%%%%%%%%%%%%%%%%%%%%%%%%%%%%%%%%
%%%%%%%%%%%%%%%%%%%%%%%%%%%%%%%%%%%%%%%%%%%%%%%%%%%%%%%%%%%%%%%%%%%%%%%%%%%
\section{Numerical Method}
\label{sec:numerical}

We next describe a numerical method for solving the governing equations
based on a cell-centered finite volume approximation.  This will be used
to validate our asymptotic results in the case of an isotropic
conductivity, and to produce comparisons for the anisotropic case.  To
simplify the discrete equations, it is helpful to first transform the
radial coordinate for the tapered annular cylindrical domain.

%%%%%%%%%%%%%%%%%%%%%%%%%%%%%%%%%%%%%%%%%%%%%%%%%%%%%%%%%%%%%%%%%%%%%%%%%%%
\subsection{Coordinate Transformation}

The radius of the tapered cylinder obeys $\gamma \Rout(z) \leqslant r
\leqslant \Rout(z)$, which suggests defining a transformed radial
coordinate $\rr=r/\Rout(z)$ that is bounded between $\gamma \leqslant
\rr \leqslant 1$ for all $0 \leqslant z \leqslant \Ht$.  The spatial
derivatives within the governing equations can then be transformed via
\begin{gather*}
  \frac{\partial}{\partial r} = \frac{1}{\Rout(z)}
  \frac{\partial}{\partial \rr} 
  \qquad \text{and} \qquad
  \frac{\partial}{\partial z} = - C \rr\,\frac{\partial}{\partial \rr} +
  \frac{\partial}{\partial z}, 
\end{gather*}
where $C \defeq \Rout^\prime/\Rout = -\alpha/\Ht$ is a constant owing to
the special exponential form~\cref{eq:Rout} assumed for $\Rout(z)$.
Applying these transformations to the velocity
components~\cref{eq:darcy-2d} yields
\begin{gather}
  v_r = -\frac{\kappa D(s)}{\Rout(z)} \, \frac{\partial s}{\partial \rr}
  \qquad \text{and} \qquad
  v_z = - \Kz(s) + \rr C D(s) \, \frac{\partial s}{\partial \rr} 
  - D(s) \, \frac{\partial s}{\partial z}, 
  \label{eq:velocities-trans}
\end{gather}
where $D(s) \defeq \Kz(s) \psi^\prime(s)$.  Here the velocities are
written in terms of a general saturation dependent hydraulic
conductivity function $\Kz(s)$, which is useful for later comparison to
the results in \cite{chuang2006porous} (see \cref{sec:simulations});
however, the same formulas extend easily to a spatially-dependent
conductivity $\Kz(r,z)$.  Making use of these velocity components, the
continuity equation~\cref{eq:Maindim} becomes
\begin{gather}
  \frac{\partial s}{\partial t} = 
  % - \nabla\cdot v =
  - \frac{1}{\Rout(z) \rr} \, \frac{\partial \left(\rr v_r
    \right)}{\partial \rr} + \rr C \, \frac{\partial v_z}{\partial \rr} -
  \frac{\partial v_z}{\partial z} .
  \label{eq:continuity-trans}
\end{gather}
The primary advantage to transforming the radial coordinate in this
manner is that the equations above are now imposed on a rectangular
computational domain in $(\rr,z)$ space, for which standard finite
difference approximations can be applied.

%%%%%%%%%%%%%%%%%%%%%%%%%%%%%%%%%%%%%%%%%%%%%%%%%%%%%%%%%%%%%%%%%%%%%%%%%%%
\subsection{Cell-centered Finite Volume Discretization}

\begin{figure}[tbhp]
  \centering\footnotesize
  \includegraphics[width=0.75\textwidth]{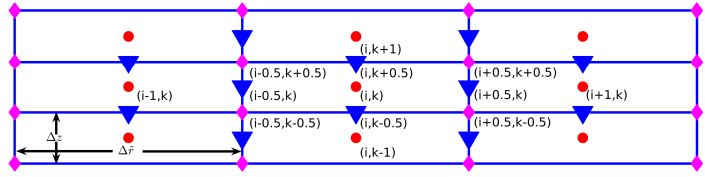}%ation}
  \caption{Discrete grid point locations in transformed coordinates,
    with points indexed as $(\rr_i, z_k)$.  Saturation is approximated
    at cell centers (red circles) and velocity components on cell edges
    (blue triangles).  The discrete equations involve saturations at
    cell corners (magenta diamonds, approximated using an average of
    cell-centered values) so that the difference equations for
    $ds_{i,k}/dt$ correspond to a full nine-point stencil involving the
    neighbouring saturation points denoted in red.}
  \label{fig:stencil}
\end{figure}

We now discretize the transformed governing equations using a
cell-centered finite volume scheme, so that we preserve as accurately as
possible the conservation of mass embodied in the continuity
equation. The computational domain is divided into an $N_r\times N_z$
rectangular grid of cells having centers
\begin{gather}
  (\rr_i, \, z_k) = \Big( \gamma + {\textstyle \left(i-\frac{1}{2}\right)}
    \Delta\rr,  \; {\textstyle \left(k-\frac{1}{2}\right)} \Delta z \Big) , 
\end{gather}
with $\Delta\rr = (1-\gamma)/N_r$ and $\Delta z=H/N_z$ for $i=1, 2,
\dots, N_r$ and $k=1, 2, \dots, N_z$.  As shown in \cref{fig:stencil},
the saturation $s_{i,k}$ is approximated at cell centers, whereas the
velocity components are located at the center of each cell edge, denoted
for example by $(v_{r,z})_{i\pm\frac{1}{2},k}$ and
$(v_{r,z})_{i,k\pm\frac{1}{2}}$.  The spatial derivatives in
\cref{eq:velocities-trans} are then replaced using compact
centered difference formulas to obtain edge-centered velocities 
\begin{align}
  (v_r)_{i+\frac{1}{2},k} &= - \left(\frac{D_{i+\frac{1}{2},k}}{\Rout_k
      \Delta \rr}\right)\left(s_{i+1,k}-s_{i k}\right), 
  \label{eq:discrete-vr} \\ 
  \left(v_z\right)_{i,k+\frac{1}{2}} &= 
  - K_{i,k+\frac{1}{2}} 
  + \left(\frac{\rr_{i} CD_{i,k+\frac{1}{2}}}{\Delta 
      \rr}\right)\left(s_{i+\frac{1}{2},k+\frac{1}{2}}-s_{i-\frac{1}{2},
      k+\frac{1}{2}}\right)
  - \left(\frac{D_{i,k+\frac{1}{2}}}{\Delta
      z}\right)\left(s_{i,k+1}-s_{i k}\right), 
  \label{eq:discrete-vz1}\\
  \left(v_z\right)_{i+\frac{1}{2},k} &= 
  - K_{i+\frac{1}{2},k} 
  + \left(\frac{\rr_{i+\frac{1}{2}} CD_{i+\frac{1}{2},k}}{\Delta
      \rr}\right)\left(s_{i+1,k}-s_{i k}\right) 
  - \left(\frac{D_{i+\frac{1}{2},k}}{\Delta
      z}\right)\left(s_{i+\frac{1}{2},k+\frac{1}{2}}-s_{i+\frac{1}{2},
      k-\frac{1}{2}}\right). 
  \label{eq:discrete-vz2}
\end{align}
Note that to maintain both stencil compactness and second-order accuracy
we have had to introduce approximate values of saturation at cell edges
and corners along with corresponding approximations for $D(s)$, all of
which are computed using appropriate arithmetic averages of
cell-centered saturation values.  The velocity components are then
substituted into the discrete continuity equation to get
\begin{multline}
  \frac{\partial s_{i,k}}{\partial t} = -\frac{1}{\rr_i \Rout_k \Delta
    \rr} \left(\rr_{i+\frac{1}{2}}(v_r)_{i+\frac{1}{2},k} -
    \rr_{i-\frac{1}{2}}(v_r)_{i-\frac{1}{2},k}\right) 
  + \frac{\rr_i C}{\Delta \rr}
  \left((v_z)_{i+\frac{1}{2},k}-(v_z)_{i-\frac{1}{2},k}\right)\\
  -\frac{1}{\Delta
    z}\left((v_z)_{i,k+\frac{1}{2}}-(v_z)_{i,k-\frac{1}{2}}\right) .
  \label{eq:discrete-s} 
\end{multline}

To discretize the boundary conditions, we set the normal velocity at the
center of each boundary cell edge to the velocity specified in the
corresponding boundary conditions \cref{eq:sapBCtop},
\cref{eq:sapBCleft} and \cref{eq:sapBCRout}. As for the remaining lower
boundary, we introduce a band of fictitious cells with centers located
one-half grid spacing below the boundary and define values of
saturation at these fictitious points. Using the boundary cell values and
the lower boundary edge value in \cref{eq:sapBCbottom}, a linear
extrapolation is used to calculate the saturation in the fictitious
cells. Complete details of the spatial discretization are provided in
\cite{janbek2017phdthesis}.  To integrate the equations in time, we use
a method-of-lines approach in which the spatially-discrete equations are
treated as a system of time-dependent ODEs, which is then integrated in
time using Matlab's stiff ODE solver {\tt ode15s} with tolerance values
$\mathtt{ABSTOL} = \mathtt{RELTOL} = 1\times 10^{-10}$.

In order to verify that our Matlab implementation yields the desired
order of spatial accuracy, we performed a numerical convergence study by
choosing a sequence of spatial grids with $N_r=N_z=32, 64, 128, 256,
512$ and computing the solution on a fixed time interval that roughly
reaches a steady state.  The error for each simulation is estimated by
treating the fine-grid solution as the ``exact solution'' and then
calculating the 1-norm difference.  The resulting errors exhibit a
convergence rate of approximately 1.99 which is strongly indicative of
second-order accuracy.

%%%%%%%%%%%%%%%%%%%%%%%%%%%%%%%%%%%%%%%%%%%%%%%%%%%%%%%%%%%%%%%%%%%%%%%%%%%
\subsection{Comparison With Experimental Data on Norway Spruce}
\label{sec:simulations}

\begin{figure}[tbhp]
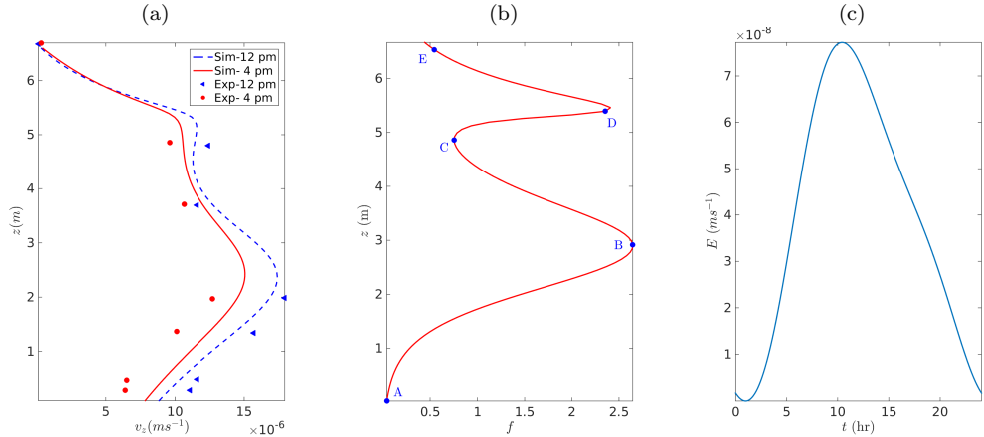

  \centering\footnotesize
  \begin{tabular}{ccc}
    (a) & (b) & (c) \\
    \includegraphics[width=0.27\textwidth,trim=0 0 0 28,clip]{newfigs2/vzExperimentalChuangComparison}
    & \includegraphics[width=0.27\textwidth,trim=0 0 0 28,clip]{newfigs2/fz}
    & \includegraphics[width=0.27\textwidth,trim=0 0 0 28,clip]{newfigs2/Et}
  \end{tabular}
  \caption[Comparison of 3D simulations with the experimental data on
  Norway spruce in~\cite{chuang2006porous} (reproduced with permission
  of Elsevier).  (a)~Vertical sap flux from simulations (Sim) and
  experiments (Exp), converted to SI units.  (b)~Height-dependent
  transpiration factor $f(z)$ from \crefrange{eq:fz}{eq:lambdaz}, where
  labels A--E are referenced later in \cref{fig:Kr_muchless_Kz}.
  (c)~Time-dependent transpiration factor $E(t)$ from
  \cref{eq:Et}.]{Comparison of 3D simulations with experimental data on
    Norway spruce.  (a)~Vertical sap flux from simulations (Sim) and
    experiments (Exp -- data extracted
    from~\cite[Fig.~6]{chuang2006porous}, with permission of Elsevier).
    (b)~Height-dependent transpiration factor $f(z)$ from
    \crefrange{eq:fz}{eq:lambdaz}, where labels A--E are referenced
    later in \cref{fig:Kr_muchless_Kz}.  (c)~Time-dependent
    transpiration factor $E(t)$ from \cref{eq:Et}.}
  \label{fig:NumericalVsData} 
\end{figure}

We now compare results from the 3D numerical scheme with the
experimental data on vertical sap flux provided by Chuang et
al.~\cite{chuang2006porous}.  For this simulation, we use parameters
listed in \cref{tab:tableValues}, except that in order to be consistent
with the 1D model we assume the entire stem is made of conductive
sapwood ($\gamma=0$) that is isotropic ($\kappa=1$).
\Cref{fig:NumericalVsData}a depicts the measured sap flux (in the
vertical direction) alongside our finite volume simulations at two
different times (noon and 4:00~pm).  The model results clearly capture
the overall solution behaviour, exhibiting the same ``double-peak''
behaviour in sap flux that can be attributed to the bimodal behaviour of
the transpiration function $f(z)$ from \crefrange{eq:fz}{eq:lambdaz}
(shown in \cref{fig:NumericalVsData}b).  In addition to this qualitative
agreement, we note that the magnitude of the sap flux is also reasonably
well-approximated by computations.  Simulations of the 1D model
equations of Chuang et al.\ were also performed using an analogous
finite volume scheme and the results were indistinguishable to the eye
from the results in \cref{fig:NumericalVsData}.

\begin{figure}[tbhp]
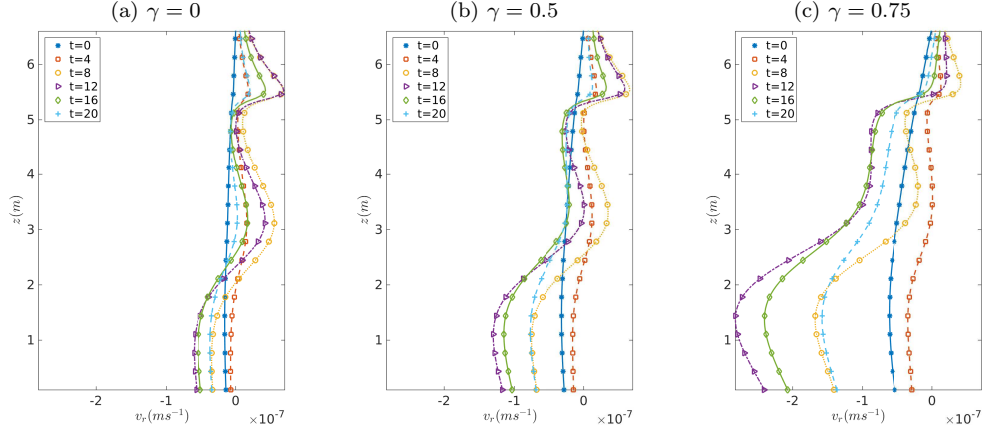

  \centering\footnotesize
  \begin{tabular}{ccc}
    (a) $\gamma=0$ & (b) $\gamma=0.5$ & (c) $\gamma=0.75$ \\
    \includegraphics[width=0.27\textwidth,trim=0 0 0 40,clip]{newfigs2/RadialVelocityComponentVerticalProfileGamma0ExpBoundaryR0pt5xlim} & 
    \includegraphics[width=0.27\textwidth,trim=0 0 0 40,clip]{newfigs2/RadialVelocityComponentVerticalProfileGamma0pt5ExpBoundaryR0pt5xlim} & 
    \includegraphics[width=0.27\textwidth,trim=0 0 0 40,clip]{newfigs2/RadialVelocityComponentVerticalProfileGamma0pt75ExpBoundaryR0pt5xlim}
  \end{tabular}
  \caption{Simulations of radial velocity $v_r$ (SI units) shown at
    various times throughout a diurnal cycle at the middle of the
    sapwood region, $r=\Rout(z)(\gamma+1)/2$.  Results are shown for
    three values of heartwood fraction $\gamma=0, 0.5, 0.75$ and 
    other parameters as in \cref{tab:tableValues} except
    $\kappa=1$.}
 \label{fig:RadialVelocityHeartwood} 
\end{figure}

The primary reasons for developing this 3D sap flow model are to capture
radial velocity (as well as radial variations in the solution) and to
investigate the impact of including a non-conducting heartwood region
with $\gamma>0$.  With this in mind, we performed a series of three
simulations with different heartwood thickness ($\gamma=0$, 0.5 and
0.75) and plotted the resulting radial velocities in
\cref{fig:RadialVelocityHeartwood}.  Two positive peaks appear in $v_r$
which clearly derive from the local maxima in the transpiration flux,
and these are offset by a comparatively large negative radial velocity
at the tree base due to root influx.  This is a geometric effect that
mimics the inward radial tilt of sapwood vessels, which due to stem
taper is largest at the base.  The effect of this radially-inward flow
is accentuated as the thickness of the annulus decreases (i.e., as
$\gamma$ increases) in order to maintain a total mass balance that
matches the specified outward transpiration flux.

%%%%%%%%%%%%%%%%%%%%%%%%%%%%%%%%%%%%%%%%%%%%%%%%%%%%%%%%%%%%%%%%%%%%%%%%%%%
%%%%%%%%%%%%%%%%%%%%%%%%%%%%%%%%%%%%%%%%%%%%%%%%%%%%%%%%%%%%%%%%%%%%%%%%%%%
%%%%%%%%%%%%%%%%%%%%%%%%%%%%%%%%%%%%%%%%%%%%%%%%%%%%%%%%%%%%%%%%%%%%%%%%%%%
%%%%%%%%%%%%%%%%%%%%%%%%%%%%%%%%%%%%%%%%%%%%%%%%%%%%%%%%%%%%%%%%%%%%%%%%%%%
%%%%%%%%%%%%%%%%%%%%%%%%%%%%%%%%%%%%%%%%%%%%%%%%%%%%%%%%%%%%%%%%%%%%%%%%%%%
\section{Asymptotic Analysis}
\label{sec:asymptotics}

The asymptotic derivation in this section bears some resemblance to
Ke\-vor\-kian and Cole's analysis of heat conduction in a long circular
rod~\cite[chapter~4]{kevorkian1981perturbation}.  Our results hold for
the general case of sap flow in an annular stem cross-section
\cref{eq:domain} as well as a saturation-dependent conductivity, and we
make no assumption on the functional form of either $\Rout(z)$ or the
transpiration flux defined via $f(z)$ and $E(t)$.  However, for the
time-varying solution we will exploit simplifications that arise from
assuming an exponential taper \cref{eq:Rout} and constant hydraulic
conductivity.  We begin by considering the special case when the radial
and axial conductivities $\Kr$ and $\Kz$ are the same order of magnitude
so that $\kappa\sim{1}$, and defer study of the anisotropic case $\Kr
\ll \Kz$ to \cref{sec:Kr_less_Kz}.

\subsection{Non-Dimensionalization in the Nearly-Saturated Regime}
\label{sec:nondim}

We next recast the 3D model in dimensionless form in order to reduce the
number of free parameters and identify important dimensionless groupings
(an analogous version of the 1D model can be derived as well, be we
don't present it here).  Introduce the following non-dimensional variables
\begin{subequations}\label{eq:Nondim}
\begin{gather}
  r = \Rmax r^*, \qquad
  z = \Ht z^*,   \qquad 
  t = {\displaystyle \frac{\tau}{2\pi}\, t^*},
  \label{eq:Nondim-rzt} \\
  \intertext{as well as rescaled versions of the stem radius and
    transpiration functions}
  \Rout(z) = \Rmax \Rout^*(z^*), \qquad
  f(z) = f_o f^*(z^*), \qquad
  E(t) = E_o E^*(t^*), 
  \label{eq:Nondim-REf} \\
  \intertext{where $f_o = 2.6$ chosen equal to the maximum value 
    of $f(z)$ so that $f^* = \bigoh{1}$.\ 
    % Base $f_o$ on the average value 1.5 instead of $f_o =
    % \frac{\ell_o}{2 \pi \Rmax} \approx 37.75$ because a further
    % scaling factor of 0.05-ish comes from 
    % \max [ \ell(z) \ \lambda(z) / \Rout(z) ].  This is evident from the
    % picture of f(z) computed with the code 'sourcefuncs.m' which
    % produces the plot in \cref{fig:fz} (below). 
    Finally, we assume that the tree remains close to a fully
    saturated state so that the saturation variable may be written}
  s(r,z,t) = \smax \left( 1 - \delta S(r^*, z^*, t^*) \right),
  \label{eq:Nondim-s}
\end{gather}
\end{subequations}
where $\delta$ measures the relative deviation from full saturation and
typically satisfies $0<\delta \ll 1$.  In practice, $\delta$ increases
with transpiration rate $E_o$, although its precise value is not so
important since $\delta$ ultimately cancels out from the final result in
our asymptotic derivation.  A rough estimate for $\delta$ can be found
by recognizing that the threshold pressure for embolism in spruce has
been measured as $\psi\approx -255\;\myunit{m}$~\cite{lu-etal-1996},
corresponding to a saturation of $s\approx 0.405$.  Because we are
interested in ``normal'' flow conditions that are well-removed from any
embolized state, we should thus constrain $\delta \lesssim 0.1$.  In
particular, using a ``base value'' of $E_o=1\times10^{-9}$, 
simulations yield a $\delta = \bigoh{10^{-2}}$ so that 
saturation is guaranteed to remain close enough to $\smax$ (further
discussion of $\delta$ is provided at the end of this section).

% * Chuang et al. (2006) have \smax = 0.5735.
% * Lu et al. (1996) indicate that embolism occurs in spruce near a
%   pressure of -2.5 MPa.
% * This corresponds to \psi = -2.5 MPa / (rho*g) = -255 m.
% * Inverting equation \cref{eq:psis} means
%       s = \smax * [ 1 - \psi/\psi_o]^{-n} \approx 0.7061 * \smax.
% * This means that saturation is within the range [0.405, 0.574].
% * Given that s ~ \smax * (1 - \delta S) and S ~ 1, this means that
%   \delta ~ 1 - s/\smax so that 0 <= \delta <= 0.3.

This rescaling has been performed such that $S$ and all ``starred''
variables can be considered $\bigoh{1}$ quantities.  Before rescaling
the governing equations, we note that the hydrostatic pressure
derivative $\psi^\prime(s)$ appears in the saturation equation
\cref{eq:Maindim} and the velocity boundary conditions through the Darcy
velocities \cref{eq:darcy-2d}.  Because $\psi$ depends nonlinearly on
saturation in \cref{eq:psis}, we will ultimately need to expand $\psi$
as a power series as well.  Therefore we perform this expansion step for
$\psi$ now, which will also permit us to determine an appropriate set of
dimensionless parameters that characterize the problem.  To this end, we
change variables using \cref{eq:Nondim} and expand $\psi$,
$\psi^\prime$ as power series in $\delta$ to obtain
\begin{subequations}\label{eq:psi-dpsi}
\begin{align}
  \psi & = - \frac{\delta \psi_o}{n}\, \left[ S + \frac{1}{2} \psi_1
    \delta S^2 + \frac{1}{6} \psi_2 \delta^2 S^3 + \bigoh{\delta^3} \right]
  \label{eq:psi},\\
  \psi^\prime &= \phantom{-} \frac{\psi_o}{n\smax} \, 
  \left[ 1 + \psi_1 \delta S + \frac{1}{2} \psi_2 \delta^2 S^2 + 
    \bigoh{\delta^3} \right], 
  \label{eq:dpsi}
\end{align}
\end{subequations}
where $\psi_1 = 1 + \frac{1}{n}$ and $\psi_2 = \left( 1+\frac{1}{n}
\right) \left( 2+\frac{1}{n} \right) = 2 + \bigoh{\frac{1}{n}}$
(recalling that $n=400$ is large).  Next, substitute the rescaled
variables \cref{eq:Nondim} into the saturation equation
\cref{eq:Maindim}, which after rearranging yields
\begin{gather*}
  \left(\frac{2\pi n\smax\Rmax^2}{\tau\psi_o K_o}\right)
  \frac{\partial S}{\partial t^*}
  = \frac{1}{r^*} \frac{\partial}{\partial r^*} \left[ \kappa K^*
    \left(1 + \psi_1\delta S\right) r^* \frac{\partial S}{\partial r^*} \right]
  + \left(\frac{\Rmax}{\Ht}\right)^2 \frac{\partial}{\partial z^*}
  \left[ K^* \left(1 + \psi_1\delta S\right) \frac{\partial S}{\partial
      z^*} \right] - \left(\frac{n\Rmax^2}{\delta\Ht\psi_o}\right) 
  \frac{\partial K^*}{\partial z^*},
\end{gather*}
noting that only the first two terms in the series \cref{eq:psi-dpsi} are
required.  Upon careful consideration of the various factors multiplying
each term, we are led to introduce the following three dimensionless
ratios
\begin{gather}
  \zeta = \frac{\Rmax}{\Ht} \approx 0.00963, \qquad 
  \eta  = \frac{2\pi n \smax \Ht^2}{\tau \psi_o K_o} \approx 4.77, \qquad 
  \mu = \frac{n\Ht}{\psi_o} \approx 0.00915.
  %\eta  = \frac{2\pi n \smax \Rmax^2}{\tau \psi_o K_o} \approx 4.42 \times 10^{-4},
  \label{eq:Nondimparam}
\end{gather}
The parameter $\zeta \ll 1$ has an obvious physical interpretation as
the stem aspect ratio and will play a central role as the primary
expansion parameter in our asymptotic analysis.  The rescaled saturation
equation may be rewritten in terms of these parameters (after dividing
by $\zeta^2$) as
\begin{gather}
  \eta \, \frac{\partial S}{\partial t}
  = \left( \frac{\kappa}{\zeta^2} \right) \frac{1}{r} 
    \frac{\partial}{\partial r} \left[ K 
    \left(1 + \psi_1\delta S\right) r\, \frac{\partial S}{\partial r} \right]
  + \, \frac{\partial}{\partial z}
  \left[ K \left(1 + \psi_1\delta S\right) \frac{\partial S}{\partial z}
  \right] - \left(\frac{\mu}{\delta}\right) \,  
  \frac{\partial K}{\partial z},
  \label{eq:scaled-sat}
\end{gather}
where all ``stars'' have been omitted on dimensionless quantities to
streamline notation.  For transpiration-driven flow we are interested only
in the case where gravitational effects are significant at leading
order, which means that the final term in \cref{eq:scaled-sat} must have
$\mu/\delta\sim{1}$.  This implies that all terms are balanced except
possibly the second, whose relative importance depends on the anisotropy
parameter through the ratio $\kappa/\zeta^2$ -- in practice, $\kappa$
varies between $1$ and $\zeta^{-1}$ although we will begin considering
the lower limit $\kappa\sim{1}$ in the asymptotics.

We next consider the boundary conditions and assume for the present that
$\gamma=0$ in \cref{eq:domain}, which corresponds to a stem consisting
entirely of conducting sapwood (i.e., no heartwood).  This does not
restrict the generality of our results since we can still capture the
effect of $\gamma>0$ by introducing an appropriate spatial variation in
$K$.  Taking $\gamma=0$ introduces a major simplification in that the
non-dimensional form of \cref{eq:sapBCleft} becomes
\begin{gather}
  \left.\frac{\partial S}{\partial r}\right|_{r=0} = 0.
  \label{eq:BCleft}
\end{gather}
The corresponding top and bottom boundary conditions
\cref{eq:sapBCtop,eq:sapBCbottom} become respectively
\begin{align}
  \left. S \right|_{z=0} = 0 %\label{eq:BCbottom}
  \qquad \text{and} \qquad
  \left. \left( 1 + \psi_1\delta S \right) \frac{\partial S}{\partial z}
  \right|_{z=1} &= \frac{\mu}{\delta}. %\label{eq:BCtop} \\[2pt]
  \label{eq:BCbottom-top}
\end{align}
It is in this second boundary condition that the earlier assumption of
$\mu\sim\delta$ ensures that the gravitational effects also contribute
to the top boundary condition at leading order.

The derivation of the transpiration flux condition \cref{eq:sapBCRout}
at the outer trunk surface is complicated by the presence of the curved
boundary where the normal direction is not aligned with a coordinate
axis.  After changing variables and expanding terms involving
$\psi^\prime$, this boundary condition reduces to
\begin{align}
  n_r \kappa K \frac{\partial S}{\partial r}
  + n_z \zeta K \left[ \frac{\partial S}{\partial z} -
    \frac{\mu}{\delta} \left(1 - \psi_1 \delta S \right) \right] =
  \frac{\phi}{2\delta} \, \zeta^2
  \left(1 - \psi_1\delta S \right) f(z) E(t) ,
  \label{eq:BCright}
\end{align}
where we have introduced the new dimensionless parameter
\begin{gather}
  \phi = \frac{2f_o E_o\mu}{K_o\zeta} \approx 0.00920 % \approx 0.105.
  \label{eq:Nondimparam-phi}
\end{gather}
which means that $\phi\sim \zeta$. Thus, based on the ``base value'' for
$E_o=1\times10^{-9}$, the primary dimensionless parameters obey the
equivalence $\delta\sim \zeta\sim \mu\sim \phi$.  We note nonetheless that the
asymptotic solution still yields an accurate approximation
for much larger values of $E_o$ that violate this equivalence, which 
we will see later in \cref{fig:AsymptoticNumericalLarge}.

Because of the small aspect ratio, the radial and vertical components of
the outward-pointing normal may be expanded as power series in small
$\zeta$:
\begin{align*}
  n_r = \frac{1}{\sqrt{1+(\zeta\Rout^\prime)^2}} 
  = 1 - \frac{1}{2} (\zeta \Rout^\prime)^2 + \dots 
  %\\
  \qquad \text{and} \qquad
  n_z = \frac{-\zeta\Rout^\prime}{\sqrt{1 + (\zeta\Rout^\prime)^2}} 
  = -\zeta\Rout^\prime + \frac{1}{2} (\zeta\Rout^\prime)^3 + \dots 
\end{align*}
These expressions are exploited in the next section to simplify the
boundary condition~\cref{eq:BCright}.

%%%%%%%%%%%%%%%%%%%%%%%%%%%%%%%%%%%%%%%%%%%%%%%%%%%%%%%%%%%%%%%%%%%%%%%%%%%
\subsection{Asymptotic Expansion: General Case}

Based on the assumption that the saturation in \cref{eq:Nondim-s} stays
close to its maximum value $\smax$, we seek a regular power series
expansion of $S$ in terms of the small parameter $\zeta$ as
\begin{gather}
  S = S_0 + S_1 \zeta + S_2 \zeta^2 + S_3 \zeta^3 + \dots 
  \label{eq:Spower}
\end{gather}
Our aim is to derive equations for the first two terms $S_0$ and $S_1$
so as to capture the effect of the nonlinearity in pressure head
\cref{eq:psi}.  We will also require two additional terms up to
$\bigoh{\zeta^3}$ (involving $S_2$ and $S_3$), which are needed for
matching purposes to obtain a closed set of equations for $S_0$ and
$S_1$.  Furthermore, we will see shortly that the two leading order
solutions are independent of $r$ so that $S_2$ is also required to
determine the leading order term in the radial sap velocity. Taking the
above expansion for saturation, the leading order equation from
\cref{eq:scaled-sat} at $\bigoh{\zeta^{-2}}$ is simply
\begin{gather*}
  \frac{1}{r} \frac{\partial}{\partial r} \left( r K\frac{\partial
      S_0}{\partial r} \right) = 0.
\end{gather*}
Because conductivity $K$ must be nonzero, it follows that $S_0$ is
independent of the radial coordinate and so $S_0 = S_0(z,t)$.  The same
equation governs $S_1$ at $\bigoh{\zeta^{-1}}$, so we conclude likewise
that $S_1 = S_1(z,t)$.  Considering the boundary condition
\cref{eq:BCright} at $r=\Rout(z)$, if transpiration is to have any
effect on saturation at the first two orders $S_0$ and $S_1$ (where the
radial derivative term vanishes) then the remaining terms should
balance.  This is equivalent to requiring $\phi \sim \delta$, which is
satisfied for the parameters in \cref{tab:tableValues} with
$E_o=1\times10^{-9}$ as discussed at the end of \cref{sec:nondim}.

The next order equation in \cref{eq:scaled-sat} involving terms at
$\bigoh{\zeta^0}$ is
\begin{gather}
  \frac{\kappa}{r} \frac{\partial}{\partial r} \left( r K \frac{\partial
      S_2}{\partial r} \right) = f_2(r,z,t) \defeq
  \eta\, \frac{\partial S_0}{\partial t} -
  \frac{\partial}{\partial z} \left(K \frac{\partial S_0}{\partial
      z}\right) + \frac{\mu}{\delta}\frac{\partial K}{\partial z},
  \label{eq:S2eq}
\end{gather}
which can be integrated in $r$ and then evaluated at $r=\Rout(z)$ to
obtain  
\begin{gather*}
  \kappa \Rout K \frac{\partial S_2}{\partial r} 
  = \int_0^\Rout r f_2(r,z) \, dr 
  = \frac{1}{2} \eta \Rout^2 \frac{\partial S_0}{\partial t} 
  - \int_0^\Rout r \, \frac{\partial}{\partial z} \left( K
    \frac{\partial S_0}{\partial z}\right) \, dr 
  + \frac{\mu}{\delta} \int_0^\Rout r \, \frac{\partial K}{\partial z} \, dr. 
\end{gather*}
Taking the same order terms arising in the transpiration boundary
condition \cref{eq:BCright} yields
\begin{gather}
  \kappa \Rout K \frac{\partial S_2}{\partial r} 
  = \Rout \Rout^\prime K \frac{\partial S_0}{\partial z} 
  - \frac{\mu}{\delta} \Rout\Rout^\prime K 
  + \frac{\phi}{2\delta} \Rout f E,
  \label{eq:BCright-order-s2}
\end{gather}
where the last term is included at this order because
$\frac{\phi}{\delta} \sim 1$.  Eliminating $S_2$ from these last
two equations leads to
\begin{gather*}
  \Rout \Rout^\prime K \frac{\partial S_0}{\partial z}
  - \frac{\mu}{\delta} \Rout \Rout^\prime K 
  + \frac{\phi}{2\delta} \Rout f E 
  = \frac{1}{2} \eta \Rout^2 \frac{\partial S_0}{\partial t} 
  - \int_0^\Rout r \, \frac{\partial}{\partial z} \left( K
    \frac{\partial S_0}{\partial z}\right)\, dr 
  + \frac{\mu}{\delta} \int_0^\Rout r \,
    \frac{\partial K}{\partial z}\, dr,
\end{gather*}
which simplifies to 
\begin{gather}
  \frac{1}{2} \eta \Rout^2 \, \frac{\partial S_0}{\partial t} 
  - \frac{\partial}{\partial z} 
  \left( G \, \frac{\partial S_0}{\partial z} \right) 
  = \frac{\phi}{2\delta} \, \Rout f E
  - \frac{\mu}{\delta} \, \frac{d G}{d z},
  \label{eq:S0eq}
\end{gather}
where we have defined
\begin{gather}
  G(z) = \int_0^{\Rout(z)} r K(r,z) \, dr.
  \label{eq:Gz}
\end{gather}
Equation~\cref{eq:S0eq} can then be solved subject to the
leading order boundary conditions from \cref{eq:BCbottom-top}:
  \begin{gather}
  S_0(0,t) = 0
  \qquad \text{and} \qquad
  \frac{\partial S_0}{\partial z}(1,t) = \frac{\mu}{\delta}.
  \label{eq:S0BCs}
\end{gather}

Next we consider the $\bigoh{\zeta}$ terms in \cref{eq:scaled-sat}
which lead to the equation
\begin{gather*}
  \frac{\kappa}{r} \frac{\partial}{\partial r} 
  \left(r K \frac{\partial S_3}{\partial r}\right)
  = \eta \, \frac{\partial S_1}{\partial t}
  - \frac{\delta\psi_1\kappa}{\zeta} \, \frac{1}{r} \, 
  \frac{\partial}{\partial r} \left(r K S_0 \, 
    \frac{\partial S_2}{\partial r}\right)
  - \frac{\partial}{\partial z} 
  \left(K \, \frac{\partial S_1}{\partial z}\right)
  - \frac{\delta\psi_1}{\zeta} \frac{\partial}{\partial z}
  \left(K S_0\, \frac{\partial S_0}{\partial z}\right), 
\end{gather*}
where we used the fact that $\delta \sim \zeta$.
Integrating from $0$ to $r$ and evaluating at $r=\Rout(z)$
yields
\begin{gather*}
  \kappa \Rout K \frac{\partial S_3}{\partial r}
  = \eta \, \frac{\partial}{\partial t} 
  \int_0^\Rout r S_1\, dr 
  - \frac{\delta\psi_1\kappa}{\zeta} \Rout K S_0 \, 
  \frac{\partial S_2}{\partial r}
  - \int_0^\Rout r \, \frac{\partial}{\partial z} 
  \left( K \, \frac{\partial S_1}{\partial z}\right) \, dr
  - \frac{\delta\psi_1}{\zeta} \int_0^\Rout r \, 
  \frac{\partial}{\partial z} \left(K S_0 \, \frac{\partial
      S_0}{\partial z}\right) \, dr, 
\end{gather*}
and a second expression for $S_3$ comes from the boundary condition at
the corresponding order in \cref{eq:BCright}
\begin{gather*}
  \kappa \Rout K \frac{\partial S_3}{\partial r}
  = \Rout \Rout^\prime K \frac{\partial S_1}{\partial z}
  + \frac{\mu\psi_1}{\zeta}\, \Rout \Rout^\prime K S_0
  - \frac{\phi\psi_1}{2\zeta}\, \Rout f E S_0.
\end{gather*}
These last two equations can be combined to eliminate $S_3$, and then
further simplified by applying \cref{eq:BCright-order-s2} to obtain
\begin{gather*}
  \eta \, \frac{\partial}{\partial t} \int_0^\Rout r S_1\, dr
  - \frac{\partial}{\partial z} \int_0^\Rout r K \frac{\partial
    S_1}{\partial z} \, dr
  - \frac{\delta\psi_1}{\zeta} \frac{\partial}{\partial z} 
  \int_0^\Rout r K S_0\frac{\partial S_0}{\partial z} \, dr = 0.
\end{gather*}
Finally, using the fact that $S_0$ and $S_1$ are independent of $r$, the
integral terms can be evaluated and the governing PDE
written in the more compact form
\begin{gather}
  \frac{1}{2} \eta \Rout^2 \, \frac{\partial S_1}{\partial t} 
  - \frac{\partial}{\partial z} 
  \left( G \, \frac{\partial S_1}{\partial z} \right)
  = \frac{\delta\psi_1}{\zeta} \, 
  \frac{\partial}{\partial z} 
  \left( G S_0 \, \frac{\partial S_0}{\partial z} \right). 
  \label{eq:S1eq}
\end{gather}
This PDE can be solved for $S_1$ provided that we impose appropriate
boundary conditions, which are obtained from the $\bigoh{\zeta}$ terms
in \cref{eq:BCbottom-top} as
\begin{gather}
  S_1(0,t) = 0
  \qquad \text{and} \qquad
  \frac{\partial S_1}{\partial z}(1,t) 
  = -\frac{\mu\psi_1}{\zeta}\, S_0(1,t).
  \label{eq:S1BCs}
\end{gather}
Note that the two terms in the top boundary condition are in balance
only if $\mu \sim \zeta$, which is satisfied to a very good
approximation. 

%%%%%%%%%%%%%%%%%%%%%%%%%%%%%%%%%%%%%%%%%%%%%%%%%%%%%%%%%%%%%%%%%%%%%%%%%%%
\subsection{Steady State Solution (Constant Transpiration Rate)}

Explicit solutions can be derived for the two leading order asymptotic
terms in the special case where the transpiration rate is constant
($E(t)\equiv 1$) and the saturation has reached a steady state.  In this
case, the $S_0$ equation \cref{eq:S0eq} can be integrated twice using
the boundary conditions \cref{eq:S0BCs} to obtain
\begin{gather}
  \bar{S}_0(z) = \frac{\mu}{\delta} \, z
  + \frac{\phi}{2\delta} \int_0^z 
  \left[ \frac{1}{G(z^\prime)} \int_{z^\prime}^1 
    \Rout(w) f(w) \, dw \right] dz^\prime. 
  \label{eq:S00}
\end{gather}
The first term in $\bar{S}_0$ represents the effect of gravity due to
changes in pressure with height.  The second term captures the net
effect of transpiration, where the expression in square brackets denotes
total transpiration flux due to branches located above height
$z^\prime$, evaluated per unit stem conductivity.  The next order
saturation correction is obtained by integrating
\crefrange{eq:S1eq}{eq:S1BCs} in a similar manner, yielding
\begin{gather} 
  \bar{S}_1(z) = -\frac{\delta\psi_1}{\zeta} \int_0^z
  \bar{S}_0(z^\prime) \frac{d \bar{S}_0(z^\prime)}{d z} \,
  dz^\prime. 
%  \frac{\mu}{\zeta}\psi_1 S_0^0(1)\int_0^z\frac{G(1)}{G(x)}dx -
%  \int_0^z\left[\frac{1}{G(x)}\int_x^1 H(w)dw\right]dx   
\label{eq:S10}
\end{gather}

It is insightful at this point to draw an analogy between the formula
for the leading order saturation $\bar{S}_0$ and electric circuit
representations commonly used to model tree sap hydraulics.  To this
end, we neglect the effects of gravity in \cref{eq:S00} and rewrite the
remaining integral term as
\begin{gather}
  \bar{S}_0(z) = \int_0^z \EuScript{R}(z^\prime) \,
  \EuScript{J}(z^\prime) \, dz^\prime 
  \label{eq:sRJ}\\
  \text{where} \qquad\qquad\qquad\qquad\qquad\qquad
  \EuScript{R}(z^\prime)
  = \frac{1}{G(z^\prime)}  
  = \left[ \int_0^{\Rout(z^\prime)} r \, K(r,z^\prime) \, dr
  \right]^{-1} \notag
 \qquad\qquad\qquad\qquad\qquad\qquad
\end{gather}
can be interpreted as an average resistance to flow within the
cross-sectional stem slice lying between $z^\prime$ and $z^\prime +
dz^\prime$.  This expression has the form of a harmonic average of
conductivities $K$, which is analogous to the formula relating
electrical resistances and conductances arranged in a series circuit.
The corresponding ``current'' is
\begin{gather*}
  \EuScript{J}(z^\prime) = \frac{\phi}{2\delta} 
  \int_{z^\prime}^1 \Rout(w) f(w) \, dw, 
\end{gather*}
which represents the total transpiration rate drawn through branches
between $z=z^\prime$ and $z=1$.  The product
$\EuScript{R}\cdot\EuScript{J}$ is integrated in \cref{eq:sRJ} along
the height of the tree in order to obtain the sap potential (or
``voltage''), which when properly scaled yields the local saturation
state.

%%%%%%%%%%%%%%%%%%%%%%%%%%%%%%%%%%%%%%%%%%%%%%%%%%%%%%%%%%%%%%%%%%%%%%%%%%%
\subsection{Steady State Solution With Constant Conductivity}

Our aim in this section is to estimate the ratio of radial
to axial sap velocity and show that the reduction in
radial-versus-axial flow is primarily due to the small stem
aspect ratio ($\zeta$) and not the conductivity ratio ($\kappa$) as one
might expect.  We have already exploited the fact that
variations in hydraulic conductivity with saturation are small by
assuming that $K(r,z)$ varies only with location.  We now
assume further that spatial variations in $K$ are likewise
small so that it is reasonable to take the conductivity function
$K\equiv 1$.

Based on this assumption, \cref{eq:S2eq} may be integrated to obtain
$S_2(r,z)$, which is the first term contributing to the radial velocity
and also the lowest-order term containing any radial dependence.  The
resulting equation at steady state obeys
\begin{gather*}
  \frac{\partial \bar{S}_2}{\partial r} = -\frac{r}{2\kappa}
  \frac{d^2 \bar{S}_0}{d z^2},  
\end{gather*}
which may be substituted into \cref{eq:darcy-2d} along with earlier
approximations and \eqref{eq:S00} to obtain a leading order expression
for the (dimensional) radial velocity 
\begin{gather}
  \bar{v}_r \approx - \left(\frac{K_o \delta\zeta}{\mu}\right)
  \frac{r}{2}\, \frac{d^2 \bar{S}_0}{dz^2}
  = - \left( \frac{K_0\phi\zeta}{2\mu} \right)
  \frac{r}{2} \frac{d}{dz} \left[ \frac{1}{G(z)} \int_z^1 \Rout(w) f(w) \,
    dw \right].   
  \label{eq:vr}
\end{gather}
Note that the conductivity ratio is absent from this expression for
$\bar{v}_r$, which explains our earlier remark that the anisotropy only
influences the radial velocity via higher order terms in the
asymptotics.

Moving on to the axial velocity, we make use of the fact that when $K$
is constant equation \cref{eq:Gz} reduces to $G(z)=\frac{1}{2}
\Rout^2(z)$, which leads to a simpler form of $\bar{S}_0$ in
\cref{eq:S00} that gives the leader order saturation derivative as
\begin{gather*}
  \frac{d \bar{S}_0}{d z} \approx \frac{\mu}{\delta} +
  \frac{\phi}{\delta \Rout^2(z)} \int_z^1 \Rout(w) f(w)\, dw.
\end{gather*}
This expression can then be substituted into the axial component of
\cref{eq:darcy-2d} and simplified to obtain the estimate
\begin{gather}
  \bar{v}_z \approx \left(\frac{K_o \phi}{2\mu}\right)
  \frac{1}{\pi\Rout^2(z)} \int_z^1 2\pi \Rout(w) f(w) \, dw.
  \label{eq:vz}
\end{gather}
The relative magnitude of the two velocity components may then be
approximated by 
\begin{gather}
  \frac{\bar{v}_r}{\bar{v}_z} = \bigoh{\zeta}, 
  \label{eq:velocitiesRatio}
\end{gather}
after dropping any $\bigoh{1}$ terms.  Consequently, the velocity ratio
at steady state depends to leading order solely on stem aspect ratio,
and furthermore the radial velocity is a factor of roughly 100 times
smaller than the vertical component.  The effect of material anisotropy
on sap flow (through small $\kappa$) has no impact at leading order,
instead entering only via higher order terms in the asymptotics.  This
is the main reason that in earlier sections we restricted the asymptotic
analysis to the case $\kappa = \bigoh{1}$, since taking $\kappa\ll 1$
only pushes the effects of the anisotropy to higher order without
impacting the leading order solution.

\begin{figure}[tbhp]
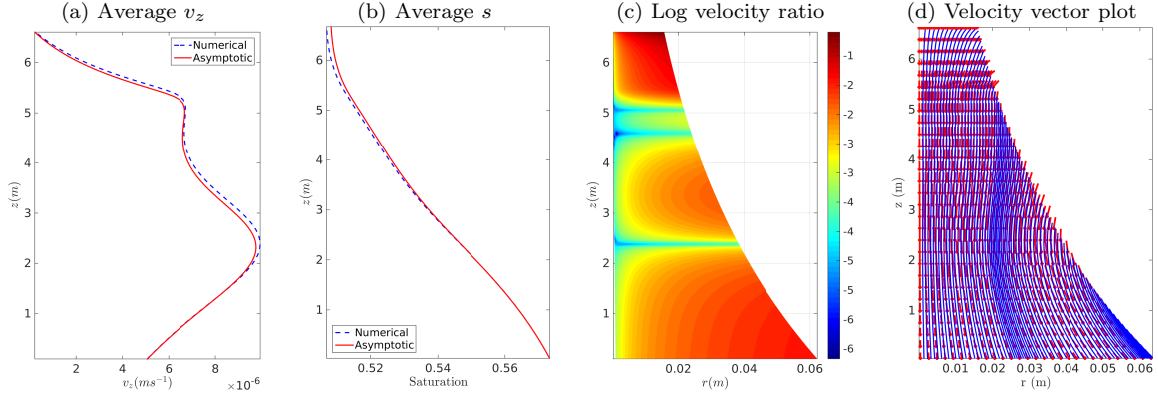

  \centering\footnotesize
  \begin{tabular}{cccc}
    (a) Average $v_z$ & (b) Average $s$ & % (c) Computed $v_r(r,z)$ & 
    (c) Log velocity ratio & (d) Velocity vector plot\\
    \includegraphics[trim=10 0 40 50,clip,width=0.222\textwidth,clip]{newfigs2/vzRadialAverageAsymNumExpBoundIsotropic}
    & \includegraphics[trim=10 0 40 50,clip,width=0.22\textwidth,clip]{newfigs2/saturationRadialAverageAsymNumExpBoundIsotropic}
    %%\imagetop{\includegraphics[trim=0 0 20 15,width=0.23\textwidth,clip]{newfigs2/vrExpBoundIsotropic}}
    & \includegraphics[trim=10 0 35 50,clip,width=0.23\textwidth,clip]{newfigs2/logRatioVrToVzExpBoundIsotropic}
    & \includegraphics[trim=10 0 40 50,clip,width=0.23\textwidth,clip]{newfigs2/StreamlinesExpBoundIsotropic}
  \end{tabular}
  \caption{Asymptotic and numerical solutions for constant transpiration
    flux and isotropic conductivity.  (a,b)~Comparison of vertical
    velocity~$v_z$ and saturation~$s$ (both averaged in radius).
    (c)~Log velocity ratio, $\log_{10}|v_r/v_z|$.  (d)~Velocity field
    arrows and streamlines.  Parameters are chosen as in
    \cref{tab:tableValues}, except that $E(t)\equiv E_o = 3.94\times
    10^{-8}$, and the asymptotic result is based on the first two terms
    in the steady-state solution.}
  \label{fig:radial_velocity}
\end{figure}

To illustrate these asymptotic results, we present in
\cref{fig:radial_velocity}a,b plots of the vertical velocity and
saturation, determined using the two leading order terms in the steady
state asymptotic solution.  In both cases, the solution variables are
averaged across the stem cross-section, and the corresponding
finite-volume numerical solution is included in order to demonstrate the
close correspondence.  The vertical velocity exhibits the characteristic
double peak that was observed for the variable-transpiration problem in
\cref{fig:NumericalVsData}.  To illustrate the relative magnitude of the
velocity components, \cref{fig:radial_velocity}c depicts the log of the
ratio $v_r/v_z$.  Except for a thin boundary layer adjacent to the top
boundary the radial component is at least a factor of 100 smaller than
the vertical component, which is consistent with our asymptotic estimate
of $\bar{v}_r/\bar{v}_z$ in \cref{eq:velocitiesRatio}.  The final plot
in \cref{fig:radial_velocity}d provides a clearer picture of the actual
flow direction within the stem by depicting both streamlines and
direction field arrows.

%%%%%%%%%%%%%%%%%%%%%%%%%%%%%%%%%%%%%%%%%%%%%%%%%%%%%%%%%%%%%%%%%%%%%%%%%%% 
\subsection{Anisotropic Conductivity With $\boldsymbol{\kappa \ll 1}$}
% $\boldsymbol{\kappa = \bigoh{\zeta^1}, \bigoh{\zeta^2}}$}
% $\boldsymbol{\Kr \ll \Kz}$}
\label{sec:Kr_less_Kz}

Our earlier assumption that the hydraulic conductivity ratio is
$\kappa=\Kr/\Kz = \bigoh{1}$ led to an asymptotic result in which the
two leading order saturations $S_0$ and $S_1$ were independent of the
degree of anisotropy in $K$.  However, sapwood is in reality a highly
anisotropic material with $10^{-4}\lesssim \kappa \lesssim
10^{-2}$~\cite{comstock-1970, redman-etal-2012}, which leads us to ask
how taking values of $\kappa \ll 1$ might alter the asymptotic solution
behaviour.  In particular, the physical values of anisotropy and
expansion parameter $\zeta=\bigoh{10^{-2}}$ suggest
considering two asymptotic limits: $\kappa=\bigoh{\zeta}$ and
$\kappa=\bigoh{\zeta^2}$.  For reasons of simplicity, we will continue
assuming that the conductivities are constant and the solution is at
steady state.

First, consider the case of moderate anisotropy with $\kappa=\zeta$,
which corresponds to taking $\Kz=K_o$ and $\Kr=\zeta K_o$ (which we note
is only an order of magnitude estimate and doesn't presume any explicit
dependence of $\kappa$ on $\zeta$).  The saturation equation
\cref{eq:scaled-sat} then becomes
% \footnote{To be completely general, we could take $\kappa=\theta\zeta$
% ($\Kr=\theta K_o\zeta$) where $\theta=\bigoh{1}$.  But I don't think
% that's necessary.}
\begin{gather*}
  %\eta \, \frac{\partial S}{\partial t} 
  0
  = \frac{1}{\zeta r} \, \frac{\partial}{\partial r} \left[ r
    \left(1+\psi_1\delta S\right) \frac{\partial S}{\partial r} \right] 
  + \frac{\partial}{\partial z} \left[ \left(1+\psi_1\delta S\right)
    \frac{\partial S}{\partial z} \right] ,
\end{gather*}
and expanding $S$ in the form of a power series~\cref{eq:Spower} yields
the leading order solution $S_0=\bar{S}_0(z)$, which is identical to
that obtained in the previous section.  However, the next order
equation for $S_1$ contains an additional term that introduces a radial
dependence of the form
\begin{gather}
  S_1(r,z) = h(z) - \left( \frac{r^2}{4} \right) 
  \frac{d^2 S_0}{dz^2},  
\end{gather}
where the function $h(z)$ is determined similarly as in the previous
section (so we omit the details here).

In the second case of a more extreme anisotropy with
$\kappa=\zeta^2$, the saturation equation~\cref{eq:scaled-sat} becomes
\begin{gather*}
  %\eta \, \frac{\partial S}{\partial t} 
  0
  = \frac{1}{r} \, \frac{\partial}{\partial r} \left[ r
    \left(1+\psi_1\delta S\right) \frac{\partial S}{\partial r} \right] 
  + \frac{\partial}{\partial z} \left[ \left(1+\psi_1\delta S\right)
    \frac{\partial S}{\partial z} \right] ,
\end{gather*}
so that the radial and vertical dependence are now fully coupled at all
orders.  Assuming for simplicity that the stem has no taper (i.e.,
$\alpha=0$ and $\Rout(z)\equiv 1$) we may transform the leading order
solution using $\Sshift_0(r,z) = S_0(r,z) - \frac{\mu z}{\delta}$, which
yields Laplace's equation $\Delta\Sshift_0 = 0$ in cylindrical
coordinates, along with boundary conditions
\begin{gather*}
  \Sshift_0 (r,0) = 0, \qquad
  \frac{\partial \Sshift_0}{\partial z}(r,1) = 0, \qquad
  \frac{\partial \Sshift_0}{\partial r}(0,z) = 0, \qquad
  \frac{\partial \Sshift_0}{\partial r}(1,z) =
  \frac{\phi}{2\delta}\, f(z) .
\end{gather*} 
The advantage of transforming $\Sshift_0$ in this manner is that the $z$
boundary conditions become homogeneous, and hence separation of
variables may be applied to obtain the series solution
\begin{gather*}
  \Sshift_0 = \sum_{n=0}^{\infty} B_n \sin(\lambda_n z) \,
  I_0(\lambda_n r) ,
  \label{eq:Kappa_ZetaSq}
\end{gather*}
where $\lambda_n = \pi(n+1/2)$, $I_0$ is the zero'th order modified
Bessel function of the first kind and 
\begin{gather*}
  B_n = \frac{\phi}{\delta\lambda_n I^\prime_0(\lambda_n)}
  \int_0^1 f(z^\prime) \sin(\lambda_n z^\prime) \; dz^\prime .
\end{gather*}
% It should be noted here that the steady state case can also be solved
% using the same method applied to the pressure head equation
% \cref{eq:Maindim}, without resort to any assumption about the magnitude
% of the saturation variation $\delta$.

\begin{figure}[tbhp]
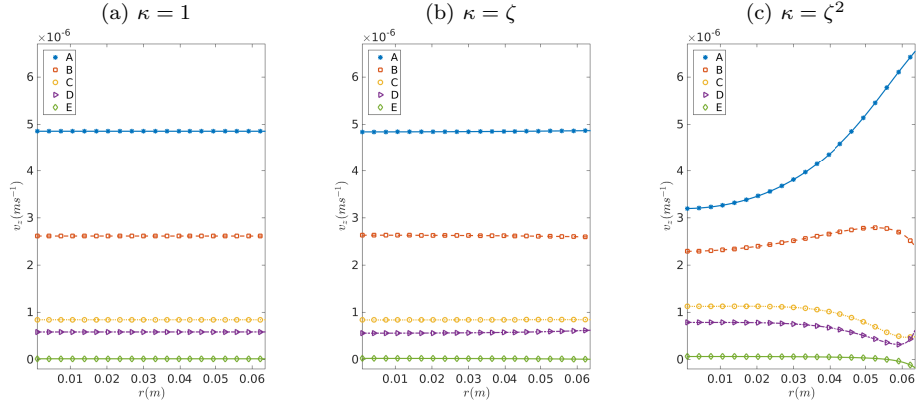

  \centering\footnotesize
  \begin{tabular}{ccc}
    %\multicolumn{1}{c}{}
    (a) $\kappa=1$ & (b) $\kappa=\zeta$ & (c) $\kappa=\zeta^2$ \\ 
    \includegraphics[trim=0 0 0 20,clip,width=0.25\textwidth]{newfigs2/vzRadialProfileCylBoundaryIsotropic}
    & \includegraphics[trim=0 0 0 20,clip,width=0.25\textwidth]{newfigs2/vzRadialProfileCylBoundaryModerateAnisotropic}
    & \includegraphics[trim=0 0 0 20,clip,width=0.25\textwidth]{newfigs2/vzRadialProfileCylBoundaryAnisotropic}
  \end{tabular}
  \caption{Effect of anisotropy ($\kappa=1$, $\zeta$, $\zeta^2$) on on
    the computed velocity in a non-tapered stem with $\alpha=0$.  The
    vertical velocity profiles are simulated numerically using a
    constant transpiration rate $E_o=3.94\times 10^{-8}$, and depicted
    at heights labelled A--E (bottom to top) on
    \cref{fig:NumericalVsData}b.}
  \label{fig:Kr_muchless_Kz} 
\end{figure}

The effect of anisotropy on the solution is investigated in
\cref{fig:Kr_muchless_Kz} where we compare the simulated vertical
velocity profiles for $\kappa=\zeta^p$, using the three exponents
$p=0,1,2$ and taking 6 terms in the Fourier--Bessel series for $p=2$.
These results are computed assuming a tree with no taper ($\alpha=0$)
and constant transpiration rate.  For each $\kappa$, we plot $v_z$ as a
function of radius at five heights corresponding to the points labelled
A--E in \cref{fig:NumericalVsData}b. For the isotropic or moderately
anisotropic cases ($p=0,1$) the velocity remains essentially constant
with radius, whereas the extreme case of $p=2$ exhibits significant
radial variations.  This is consistent with our asymptotic results which
show that radial dependence only enters the leading order solution when
$\kappa=\bigoh{\zeta^2}$, and may help to explain the radial dependence
in velocity that was observed experimentally in \cite{james2003axial,
  poyatos-etal-2007}.

\begin{figure}[tbhp]
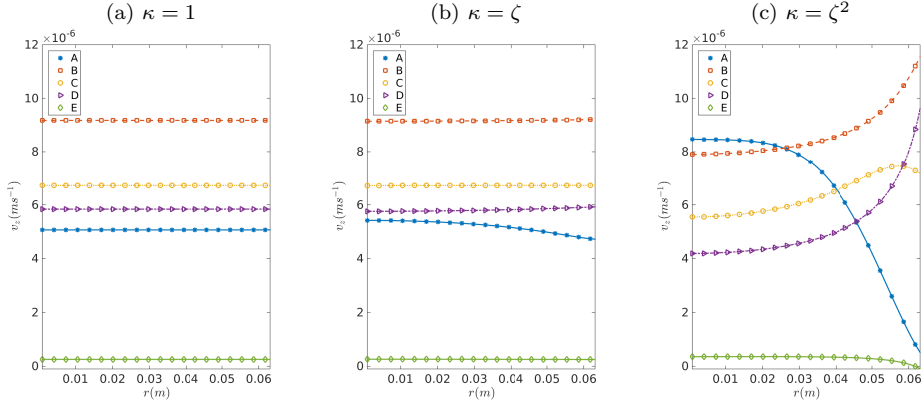

  \centering\footnotesize
  \begin{tabular}{ccc}
    %\multicolumn{1}{c}{}
    (a) $\kappa=1$ & (b) $\kappa=\zeta$ & (c) $\kappa=\zeta^2$ \\ 
    \includegraphics[trim=0 0 0 20,clip,width=0.25\textwidth]{newfigs2/vzRadialProfileExpBoundaryIsotropic}
    & \includegraphics[trim=0 0 0 20,clip,width=0.25\textwidth]{newfigs2/vzRadialProfileExpBoundaryModerateAnisotropic}
    & \includegraphics[trim=0 0 0 20,clip,width=0.25\textwidth]{newfigs2/vzRadialProfileExpBoundaryAnisotropic}
  \end{tabular}
  \caption{Same as \cref{fig:Kr_muchless_Kz} for a tapered stem with
    $\alpha=1.42$.}
  \label{fig:Kr_muchless_Kz_Exp} 
\end{figure}

This sequence of simulations was then repeated for a tapered stem with
$\alpha=1.42$ and the corresponding velocity plots are shown in
\cref{fig:Kr_muchless_Kz_Exp}. We observe similar behaviour to the
previous cases except that the $\kappa=\zeta^2$ results have a more
pronounced radial variation.  Even for the moderately anisotropic case
($\kappa=\zeta$), there is a slight radial dependence visible in the
bottom-most $v_z$ plot (location A).  It is also interesting to note
that introducing stem taper causes a significant drop
in vertical velocity near the tree base owing to the increase in sapwood
cross-section there; this should be comtrasted with the untapered case
where the vertical velocity increases monotonically with height.

We performed one further validation of in the extreme case
$\kappa=\zeta^2$ by comparing a numerical simulation with the 6-term series
solution \cref{eq:Kappa_ZetaSq}.  The relative
difference in the saturation deficit $\smax-s=\delta S$ between the
asymptotic and numerical solutions is of order $\bigoh{10^{-2}}$, which
is the same order as the missing correction term in the asymptotic
expansion of $S$, as expected.

%%%%%%%%%%%%%%%%%%%%%%%%%%%%%%%%%%%%%%%%%%%%%%%%%%%%%%%%%%%%%%%%%%%%%%%%%%% 
\subsection{Time-Dependent Transpiration Source}
\label{sec:special-case}

Next, consider a more general time-de\-pen\-dent source where the
transpiration rate $E(t)$ is expanded as a Fourier series
\begin{gather}
  E(t) = \Re \left[ \sum_{m=0}^\infty d_m \exp({\iimag m t}) \right], 
  \label{eq:Efourier}
\end{gather}
where $d_0=1$ and the remaining $d_m$ are the complex Fourier
coefficients.  In order to obtain a closed-form solution, we continue to
exploit simplifications that arise from assuming a constant, isotropic
conductivity ($\Kr=\Kz=1$ in dimensionless variables).  We also make
explicit use of the exponential form $\Rout(z)=\exp(-\alpha z)$ for the
stem taper function, so that equation \cref{eq:Gz} yields simplified
formulas for both $G(z)=\frac{1}{2}\Rout(z)^2$ and $G^\prime(z)=-\alpha
\Rout(z)^2$.  The leading order saturation $S_0$ is then taken to have
an analogous Fourier mode decomposition
\begin{gather}
  S_0(z,t) = \Re \left[ \sum_{m=0}^\infty d_m S_0^m(z) \exp({\iimag mt})  
  \right],  
  \label{eq:S0fourier}
\end{gather}
where $S_0^m(z)$ are unknown functions and the transpiration
coefficients $d_m$ are introduced as scaling factors to simplify later
expressions.  Substituting these two series into the leading order
equation \cref{eq:S0eq} and collecting terms in $\exp({\iimag mt})$
yields a sequence of linear ODEs for $S_0^m(z)$
\begin{gather}
  \frac{d^2 S_0^m}{d z^2} - 2\alpha \frac{d S_0^m}{d z} - \iimag m \eta  
  S_0^m = H_0^m(z),  
  \label{eq:S0mOrigin}
\end{gather}
where
\begin{gather}
  H_0^m(z) = -\frac{2}{\delta} \left[ \frac{\phi f(z)}{2\Rout(z)}  
    + \left\{ \begin{array}{@{\,}ll@{}}
        \alpha\mu, & m=0 \\
        0        , & m>0
      \end{array} \right\}
  \right].
  \label{eq:H0mOrigin}
\end{gather}
The corresponding boundary conditions for $S_0^m$ are obtained from
\cref{eq:S0BCs} as
\begin{gather}
  S_0^m(0) = 0
  \qquad \text{and} \qquad 
  \frac{d S_0^m(1)}{dz} = 
  \begin{cases}
    \frac{\mu}{\delta}, & m=0 \\
    0                 , & m>0
  \end{cases} .
  \label{eq:S0mBCs}
\end{gather}

The leading order term $S_0^0(z)$ has already been determined as the
steady solution $\bar{S}_0(z)$ in \cref{eq:S00}.  For the remaining ODEs
with $m\geqslant 1$, we split the general solution of
\cref{eq:S0mOrigin} into the sum of homogeneous and particular solutions 
as
\begin{subequations}
\label{eq:S0mAll}
\begin{gather}
  S_0^m = S_{0,h}^m + S_{0,p}^m,
  \label{eq:S0mboth}
\end{gather}
where the homogeneous part is
\begin{gather}
  S_{0,h}^m = A_0^{m+} \exp({\varrho_m^+ z}) + A_0^{m-} \exp({\varrho_m^-z}) , 
  \label{eq:S0mh}
\end{gather}
for constants $A_0^{m\pm}$ and
\begin{gather}
  \varrho_m^\pm = \alpha \pm \sqrt{\alpha^2 +\iimag m\eta}
  % = |\varrho_m| \exp({\iimag\varphi_m})
  \, . \label{eq:rho_m}
\end{gather}
Variation of parameters then gives the particular solution
\begin{gather}
  S_{0,p}^m(z) = 
  \frac{\exp(\varrho_m^+ z)}{\varrho_m^+ - \varrho_m^-} 
  \int_0^z \exp(-\varrho_m^+ z') H_0^m(z') \, dz'
  - \frac{\exp(\varrho_m^- z)}{\varrho_m^+ - \varrho_m^-} 
  \int_0^z \exp(-\varrho_m^- z') H_0^m(z') \, dz' , 
  \label{eq:S0mp}
\end{gather}
after which the boundary conditions \cref{eq:S0mBCs} can be substituted
into \cref{eq:S0mboth,eq:S0mh,eq:S0mp} to determine the constants 
\begin{gather}
  A_0^{m-} = -A_0^{m+} = \frac{\displaystyle
    \int_0^1 \Big[ 
    \varrho_m^+ \exp\left({\varrho_m^+(1-z')}\right) - 
    \varrho_m^- \exp\left({\varrho_m^-(1-z')}\right) \Big] \, 
    H_0^m(z') \, dz'
  }{
    (\varrho_m^+ - \varrho_m^-) \big[\varrho_m^+ \exp({\varrho_m^+}) -
    \varrho_m^- \exp({\varrho_m^-}) \big]} . 
  \label{eq:A0m}
\end{gather}
\end{subequations}

\leavethisout{
  Now 
  \begin{align}\label{eq:S0h}
    \Re\left[d_m\left(S_0^m\right)^h e^{\iimag mt}\right] = |d_m||A_0^1| &
    \left(e^{(a+|P_m|\cos(\phi_{P_m}))z} \cos\left[|P_m|\sin(\phi_{P_m})z +
        mt+\phi_{A_0^1}+\phi_{d_m}\right]\right.\\
    \notag 
    &\left.-e^{\left(a-|P_m|\cos(\phi_{P_m})\right)z}
      \cos\left[|P_m|\sin(\phi_{P_m})z-mt-\phi_{A_0^1}-\phi_{d_m}\right]\right) 
  \end{align}
  where
  \begin{gather}
    A_0^1=|A_0^1|e^{\phi_{A_0^1}} \qquad\qquad
    d_m=|d_m|e^{\phi_{d_m}}
    \label{eq:A01dm}
  \end{gather}
  Similarly
  \begin{align}\label{eq:S0p}
    &\Re\left[d_m\left(S_0^m\right)^p e^{\iimag mt}\right]=\\ \notag
    &\frac{|d_m|}{2|P_m|}\int_0^z
    H_0^m(w)\left(e^{(a+|P_m|\cos(\phi_{P_m}))(z-w)}\cos\left[|P_m|\sin(\phi_{P_m})(z-w)+mt-\phi_{P_m}+\phi_{d_m}\right]\right.\\
    \notag
    &\left.-e^{(a-|P_m|\cos(\phi_{P_m}))(z-w)}\cos\left[|P_m|\sin(\phi_{P_m})(z-w)-mt+\phi_{P_m}-\phi_{d_m}\right]\right)dw
  \end{align}
}

Proceeding to the next order in the asymptotic solution for $S_1$, a
similar series expansion
\begin{gather}
  S_1(z,t) = \Re \left[ \sum_{m=0}^\infty d_m S_1^m(z) \exp({\iimag mt})
  \right]
  \label{eq:S1fourier}
\end{gather}
is substituted into \cref{eq:S1eq}, and \cref{eq:S0eq} is used to
simplify the right hand side involving $S_0$.  The resulting equation
involves three extra nonlinear terms that require individual Fourier
series expansions:
\begin{gather*}
  \left(S_0\right)^2 = \Re \left[\sum_{m=0}^\infty d_m B_m(z) 
    e^{\iimag m t}\right], 
  \quad 
  \left(\frac{\partial S_0}{\partial z}\right)^2 = \Re
  \left[\sum_{m=0}^\infty d_m C_m(z) e^{\iimag m t}\right], 
  \quad 
  S_0 E(t) = \Re \left[\sum_{m=0}^\infty d_m D_m(z) 
    e^{\iimag m t}\right].
\end{gather*}
After some further simplification, the ODEs for $S_1^m(z)$ can be written as
\begin{gather*}
 \frac{d^2 S_1^m}{dz^2} - 2\alpha \frac{d S_1^m}{dz} - 
 \iimag m \eta S_1^m = H_1^m(z), 
  \label{eq:S1mOrigin}
\end{gather*}
which are identical to the $S_0^m(z)$ equations except that the right
hand side is given by \cref{eq:S0eq} as
\begin{gather*}
  H_1^m(z) = \frac{\psi_1\delta}{\zeta} \, \left[
    \frac{2\alpha\mu}{\delta} S_0^m(z)
    - \frac{\iimag m \eta}{2} B_m(z) - C_m(z) 
    + \frac{\phi f(z)}{\delta\Rout(z)} \, D_m(z) \right]. 
  \label{eq:H1mOrigin}
\end{gather*}
The corresponding boundary conditions from \cref{eq:S1BCs} are
\begin{gather*}
  S_1^m(0) = 0
  \qquad \text{and} \qquad
  \frac{\partial S_1^m}{\partial z}(1) 
  = -\frac{\mu\psi_1}{\zeta}\, S_0^m(1).
\end{gather*}
We have already obtained the first term ($m=0$) in the $S_1$-series as
\cref{eq:S10} from the steady state solution, while for $m\geqslant 1$
we proceed as before by splitting
\begin{gather*}
  S_1^m = S_{1,h}^m + S_{1,p}^m,
  \label{eq:S1mboth}
\end{gather*}
where the homogeneous solution is
\begin{gather*}
  S_{1,h}^m = A_1^{m+} \exp(\varrho_m^+ z) + A_1^{m-} \exp(\varrho_m^- z),
  \label{eq:S1mh}
\end{gather*}
and the particular solution $S_{1,p}^m$ is identical to \cref{eq:S0mp}
with $H_0^m$ replaced by $H_1^m$.  Finally, applying the boundary
conditions yields the coefficients
\begin{gather*}
  A_1^{m-} = -A_1^{m+} = \frac{\displaystyle
    \frac{\mu\psi_1}{\zeta} S_0^m(1)  
    + \frac{1}{\varrho_m^+ - \varrho_m^-} \int_0^1 \Big[ 
    \varrho_m^+ \exp\left({\varrho_m^+(1-z')}\right) - 
    \varrho_m^- \exp\left({\varrho_m^-(1-z')}\right) \Big] \, 
    H_1^m(z') \, dz'
  }{
    \varrho_m^+ \exp({\varrho_m^+}) - \varrho_m^- \exp({\varrho_m^-}) 
  } . 
  \label{eq:A1m}
\end{gather*}

\begin{figure}[tbhp]
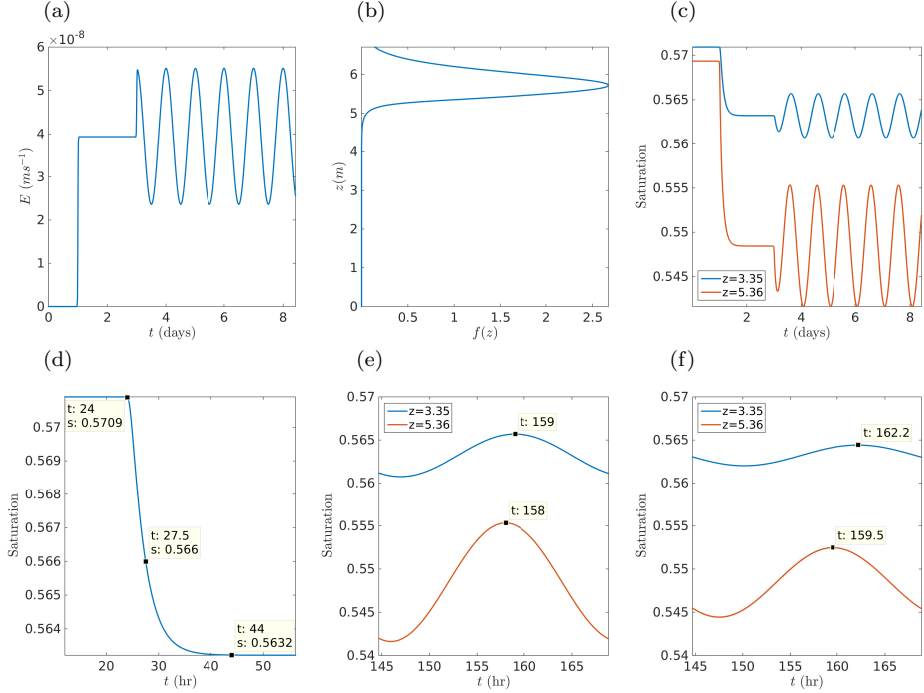

  \centering\footnotesize
  \begin{tabular}{lll}
    %\multicolumn{1}{c}{}
    (a) & (b) & (c) \\
    \includegraphics[trim=2cm 0 0 0,width=0.24\textwidth]{newfigs2/SpatialShiftRelaxationTimeTranspirationFunction} &
    \includegraphics[trim=2cm 0 0 0,width=0.24\textwidth]{newfigs2/fzUpperLeaves} &
    \includegraphics[trim=2cm 0 0 0,width=0.24\textwidth]{newfigs2/SpatialShiftRelaxationTime}\\
    (d) & (e) & (f) \\
    \includegraphics[trim=2cm 0 0 0,width=0.24\textwidth]{newfigs2/RelaxationTime3pt35} &
    \includegraphics[trim=2cm 0 0 0,width=0.24\textwidth]{newfigs2/SpatialShift} &
    \includegraphics[trim=2cm 0 0 0,width=0.24\textwidth]{newfigs2/SpatialShiftLargerEta}\\
  \end{tabular}
  \caption{Study of time scales in the asymptotic solution based on
    \cref{eq:Trelax} and \cref{eq:WaveSpeed}, with comparisons to
    numerical simulations.  For a given transient disturbance in 
    transpiration flux, simulations are shown at two 
    heights $z=0.5H, 0.8H$, with parameters $E_o=3.94\times
    10^{-8}$, $\kappa=1$, $\alpha=1.42$, $\eta=4.77$.  
    (a) Transient disturbance $E(t)$ corresponding to an
    extreme weather event.
    (b) Spatial transpiration factor $f(z)$ corresponding to a tree with
    branches concentrated in the crown. 
    % (b) Estimated time taken for travel along the entire stem
    % length versus different $\eta$ values using \cref{eq:WaveSpeed}. 
    %
    (c) Computed saturation profiles.
    (d) Zoom of (c) near the jump in the $z=0.5H$ profile. 
    (e) Zoom of (c) showing the time shift between corresponding
    travelling wave peaks at the two different heights.  
    (f) Same as (e) except using a larger value of $\eta=14.3$ obtained
    by scaling $r_0, H$, $\psi_o$ by a factor of $3$.}
  \label{fig:Spatial_Time_Shifts} 
\end{figure}

Although the formulas for $S_0^m$ and $S_1^m$ are somewhat complex, some
insights can be drawn about the behaviour of solutions by concentrating
on the underlying structure.  It is clear from \cref{eq:S0mAll} that the
leading order solution involves terms of the form $\Re [
\exp(\varrho_m^\pm z) \exp(\iimag m t) ]$, which when summed
give rise to upward- and downward-travelling saturation waves moving at
speed 
% \begin{gather}
%   \left|\frac{dz}{dt}\right| \propto
%   \frac{2m}{|\varrho_m^+-\varrho_m^-|} = \frac{m}{(\alpha^4 +
%     m^2\eta^2)^{1/4}},
%   \label{eq:WaveSpeed}
% \end{gather}
\begin{gather}
  \left|\frac{dz}{dt}\right|=\frac{2m}{\Im\left[\varrho_m^+-\varrho_m^-\right]},
  \label{eq:WaveSpeed}
\end{gather}
and with amplitude that decays with $z$ in the direction of travel.
Furthermore, in the simple case of zero forcing (i.e., no
transpiration and $E_o=0=\phi$) we can show that the leading
order term in \cref{eq:S0eq} undergoes a simple exponential decay
process from the initial state over a relaxation time scale 
\begin{gather}
  T_r = \frac{\eta}{\alpha^2 + \left(\frac{\pi}{2}\right)^2}.
  \label{eq:Trelax}
\end{gather}
In order to test these observations on a concrete example, we consider a
special time-varying transpiration source term $E(t)$ as pictured in
\cref{fig:Spatial_Time_Shifts}a that begins at zero, jumps suddenly to a
relatively large non-zero constant value (here $E_o$), and then after
some delay returns to a ``normal'' diurnal periodic cycle (a pure $m=1$
mode).  This source may be viewed physically as arising
from a severe weather event or other rapid change in ambient conditions.
To allow clear travelling waves to develop along the stem,
we also concentrate the transpiration source term in a small region near
the tree-top using the function $f(z)$ as shown in
\cref{fig:Spatial_Time_Shifts}b.

Numerical simulations yield the time-dependent saturation profiles in
\cref{fig:Spatial_Time_Shifts}c at two different stem heights.  In
response to the initial transpiration jump, there is a clear relaxation
phase where saturation decays gradually back to a constant steady state.
The zoomed plot in \cref{fig:Spatial_Time_Shifts}d shows that the
relaxation time is roughly 3.5~h, which compares well with the estimate
of $T_r\approx 4.06$~h obtained using \cref{eq:Trelax}, converted to
dimensional time.  Moving next to the diurnal variations in
transpiration, the zoomed plot of the two saturation profiles in
\cref{fig:Spatial_Time_Shifts}e demonstrates the existence of travelling
waves of saturation moving down the stem, shown as a time shift between
the two corresponding peaks.  The shift can be estimated as roughly
$1$~h, whereas the asymptotic wave speed formula \cref{eq:WaveSpeed} can
be used to estimate a shift of $1.44$~h.  A second simulation is shown
for a larger value of $\eta=14.3$ in \cref{fig:Spatial_Time_Shifts}f,
which exhibits a computed shift of 2.7~h as compared with the asymptotic
estimate of 2.86~h, which is a significantly better agreement.

\begin{figure}[tbhp]
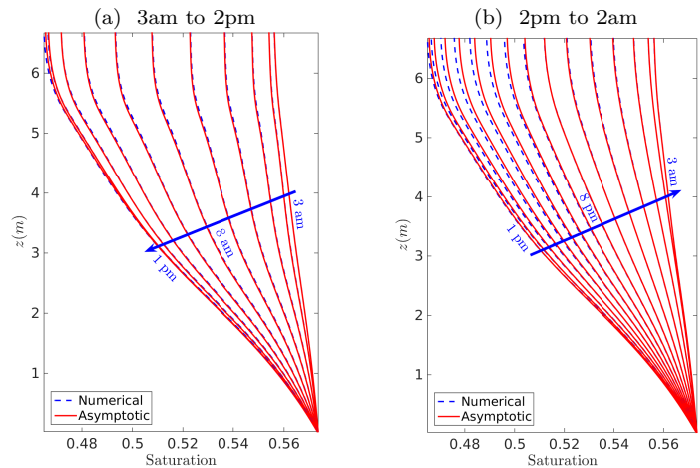

  \centering\footnotesize
  \begin{tabular}{cc}
    (a)~~3am to 2pm & (b)~~2pm to 2am\\
    \includegraphics[width=0.3\textwidth,trim=0 0 0 50,clip]{newfigs2/saturationExpBoundaryIsotropicTimeVariation2} 
    & \includegraphics[width=0.296\textwidth,trim=0 0 0 50,clip]{newfigs2/saturationExpBoundaryIsotropicTimeVariation1}
  \end{tabular}
  \caption{Saturation profiles over a one-day period, showing 
    the horizontally-averaged numerical solution and 
    asymptotic approximation.  For purposes of clarity the profiles over
    the daily cycle are separated into two periods (a,b),
    since the saturation decreases over the first half day after which it
    increases again. Parameters are $\alpha=1.42$ and $E_o=3.94\times 10^{-8}$.}
    % $\alpha=1.42$ and $E_o = 7.5\times10^{-8}$,
%     Notice that we have imposed a relatively large saturation difference
%     here (WHAT IS IT??? well outside the assumed limit of
%     $\delta=\bigoh{10^{-2}}$) in order that the difference between the
%     asymptotic/numerical solutions is visible to the eye; nevertheless,
%    the two-term asymptotic estimate provides an accurate approximation
%    even in this extreme case.}
  \label{fig:AsymptoticNumericalLarge} 
\end{figure}

One further verification of the asymptotic solution is now performed in
which we compare the radially-averaged saturation profile from
simulations to that obtained from the two-term asymptotic expansion over
a one-day period.  Taking $E_o=1\times 10^{-9}$ so that
$\mu\sim\phi\sim\zeta$, the relative difference between numerical and
asymptotic values of the saturation deficit is $\bigoh{10^{-4}}$.  It is
only when the transpiration rate is increased to $E_o=3.94\times
10^{-8}$ that we begin to move out of the asymptotic regime and
differences in saturation become visible to the naked eye as shown in
\cref{fig:AsymptoticNumericalLarge}.
  
\subsection{Discussion: Physical Relevance of Dimensionless Ratios}

To conclude this section, we highlight four dimensionless
parameters (or ratios) that play a prominent role in the asymptotic
solution just derived, and which also have straightforward physical
interpretations.

\paragraph*{Stem aspect ratio, $\zeta$} 
This parameter plays a central role in the asymptotic analysis as the
power series expansion parameter.  One of our main conclusions is that
the ratio of the radial and vertical velocity components is
${\bar{v}_{r}}/{\bar{v}_{z}} = \bigoh{\zeta}$, which is small regardless
of the conductivity ratio $\kappa$.  Indeed, this is what allows us to
assume $\kappa=1$ in our asymptotic derivation and still obtain a
solution whose leading order term is relevant to the anisotropic case.

% One dimensional parameter is worth mentioning here, which is $K_o$, as
% this is equal to $v_{z0}$ (see \cref{eq:vz}).

\paragraph*{Transpiration--flux ratio, $\chi=\phi/\delta$} 
This ratio appears in several key places throughout the asymptotic
derivation wherever transpiration terms appear in the solution,
including \cref{eq:BCright,eq:S00,eq:S0mOrigin}.
Substituting the parameter definitions into $\chi$ yields
\begin{gather*}
  \chi = \frac{2f_o E_o \mu}{K_o \zeta \delta} 
  % = \frac{f_o E_o n H^2}{K_o \psi_o r_o \delta} 
  = \frac{(2\pi r_o H) (f_o E_o)}{(\pi r_o^2) 
    \left(K_o \delta \frac{\psi_o}{nH}\right)},  
\end{gather*}
and based on the right-most expression $\chi$ may be interpreted as the
ratio of transpiration flux through the stem surface to 
vertical sap flux through a circular stem cross-section.  For the
parameters of interest in this study $\chi\sim{1}$, which reflects the
balance that must exist between these two fluxes under ``normal''
daytime conditions.  Other limits could nonetheless be considered, such
as $\chi \ll 1$ for which the transpiration rate is insufficient to
generate an appreciable change in saturation and hence the impact of
transpiration will only be felt in higher order terms.  On the other
hand, imposing a higher transpiration rate with $\chi \gg 1$ could be
viewed as shifting the tree into an embolism regime for which
saturation is no longer a smooth function, violating a fundamental
assumption in our model.

% or if we define the leaf specific
% conductivity (LSC) \cite{sterck2008persisting} as
% \begin{gather}
%   LSC=\frac{\mbox{Maximum tree conductance}}{\mbox{maximum leaf area}}
% \end{gather}
% then 
% \begin{gather}
%   \chi=\frac{\mbox{Maximum transpiration flux}}{\mbox{LSC}} = \bigoh{1}
% \end{gather}

\paragraph*{Gravity--saturation ratio, $\xi=\mu/\delta$}
This ratio also appears in the governing equations
\cref{eq:scaled-sat,eq:S0eq} and the leading order saturation boundary
conditions \cref{eq:BCbottom-top,eq:S0BCs}.  In terms of dimensional
parameters, $\xi = \frac{nH}{\psi_o\delta}$,
% \begin{gather*}
%   =\frac{\mbox{gravity force}}{\mbox{saturation dependent force}} =
%   \bigoh{1}
%   \\
%   \intertext{Note that this can be written as} 
%   \xi = \frac{ \left( \frac{\rho g \ell^2}{\sigma} \right) 
%     \left( \frac{\kappa}{\ell^2} \right) }{
%     \left( \frac{\kappa \rho g \psi_o}{n\sigma H} \right) \delta}
%   = \frac{\text{\em Bo}\, \text{\em Da}}{\text{\em Ca}\, \delta}
%   \intertext{where $\ell$ is a typical pore length scale,
%     $\kappa=\frac{K_o\bar{\mu}}{\rho g}$ is permeability, $\bar{\mu}$ is
%     viscosity, $g$ is gravitational acceleration, $\sigma$ is surface
%     tension, \emph{Bo} is Bond number, \emph{Da} is Darcy number, and
%     \emph{Ca} is capillary number.} 
% \end{gather*}
which can be viewed as a balance between the driving force due to
gravity, and the corresponding (saturation-dependent) capillary forces
acting on the pore scale in both stem and roots.  At night when
transpiration is a minimum these two forces must be in balance to
prevent water loss into the soil, which is reflected in the fact that
$\xi\sim{1}$.  Recall that our analysis requires $\mu \sim \zeta$
(when $\psi_1 \sim 1$, see \cref{eq:S1BCs}) which places a restriction
on the model parameters.  For example, if tree height and radius are
scaled up by the same factor so that $\zeta$ remains fixed then the
ratio $\psi_o/n$ must also increase, meaning that larger trees may
develop larger tensions for a given saturation deficit.

% This means that tall trees (and hence large $\mu$) will have a larger
% difference in saturation between the top and the bottom of the tree,
% even without transpiration; thus the upper parts of the tree need to
% be more resistant to embolism; otherwise, taking into account that
% $\mu = \bigoh{\zeta}$, which means that $\Ht =
% \bigoh{\zeta\bar{\psi}_0}$, the more probable explanation would be
% that taller trees have higher $\bar{\psi}_0$ value, so that small
% changes in saturation, lead to large pressure changes, and thus larger
% flow speeds.
  
\paragraph*{Time parameter, $\eta$} 
Consider the formulas for the travelling wave speed \cref{eq:WaveSpeed}
and relaxation time \cref{eq:Trelax} derived in the previous section.
For simplicity, consider a tree with no taper ($\alpha=0$) in which case
the wave speed formula reduces to $(2m/\eta)^{1/2}$ and the relaxation
time to $T_r=4\eta/\pi^2$.  Clearly, the parameter $\eta$ is intimately
tied to the time variation of the solution both through the speed of
propagation of saturation disturbances along the stem (with
characteristic time proportional to $\eta^{1/2}$) and the time for decay
of disturbances (proportional to $\eta$).

\section{Conclusions}
\label{sec:conclusions}

We have extended and generalized the 1D porous medium model for
tran\-spir\-a\-tion-driven tree sap flow developed by Chuang et
al. in~\cite{chuang2006porous} to a tapered 3D axisymmetric stem
geometry.  Methods of asymptotic analysis are used to derive the first
two terms in a regular asymptotic series expansion in powers of the
aspect ratio $\zeta$, which are then used to obtain formulas for the
spatial and temporal variations of the saturation and velocity
components.  Various possible flow regimes are studied through the use
of several dimensionless ratios.  The results are illustrated using a
set of physical parameters and nonlinear coefficient functions
corresponding to Norway spruce trees, although the analysis applies to a
much more general class of parameters.  One interesting conclusion of
our analysis is that moderate levels of anisotropy $\kappa$ in the
hydraulic conductivity do not induce any radial flow (or radial solution
dependence) at leading order so that the simpler isotropic porous medium
solution is a reasonable approximation for moderate $\kappa$.  Indeed,
introducing an anisotropy $\kappa\sim\zeta^p$ that depends on powers
$p=0,1,2$ has the effect of successively ``shifting'' radial variations
up to the higher order terms in the asymptotic solution.

The asymptotic results are verified using a second order finite
volume approximation of the original governing equations, showing the
results to be accurate for a relatively large range of saturations (as
long as we avoid the low pressure regime where embolisms are likely to
form, at which point the model assumptions are no longer valid).  An
interesting correspondence is drawn between asymptotic results for
the steady state case and the more pervasive circuit model representation
for tree sap flow.  Furthermore, the vertical mass flux was shown to
agree with experimental results from \cite{chuang2006porous}.

In future, we plan to extend the model to a more general non-symmetric
3D geometry where the solution either experiences angular variations or
a more complicated branching distribution along the stem.  This model
also forms an ideal platform from which to study the interplay between
transpiration and embolism formation under more extreme conditions.

Our asymptotic analysis will also facilitate the study of inverse
problems related to estimating the transpiration functions $f(z)$ and
$E(t)$, as well as model parameters such as $\mu$, $\eta$ and $\phi$.
Given a set of noisy measurements of sap velocity, we can use our
formulas to recover estimates of $f(z)$ and $E(t)$. Then if $\eta$ is
small enough that saturation relaxes to the zero transpiration
steady-state at night, we may estimate $\mu$ using \cref{eq:S00}. The
parameter $\eta$ can then be determined using measurements of the
relaxation time at night in \cref{eq:Trelax}. Finally, assuming that
data for $K_o$ is available from vulnerability curve measurements,
$\phi$ can be estimated making use of values for $f_o$ and $E_o$. The
details of this estimation procedure can be found in
\cite{janbek2017phdthesis}.

\bibliographystyle{siam}
\bibliography{SapFlowPaper21}

\end{document}